\documentclass{emulateapj}

\newcommand{\msun}{\mbox{$M_{\odot}$}}
\newcommand{\lsun}{\mbox{$L_{\odot}$}}
\newcommand{\zsun}{\mbox{$Z_{\odot}$}}
\newcommand{\lx}{\mbox{$L_X$}}
\newcommand{\tx}{\mbox{$T_X$}}
\newcommand{\lb}{\mbox{$L_B$}}

\shorttitle{X-Rays from Hot Gas in Mergers}
\shortauthors{Cox et al.}

\begin{document}

\title{X-Ray Emission from Hot Gas in Galaxy Mergers}

\author{T. J. Cox\altaffilmark{1},
Tiziana Di Matteo\altaffilmark{2},
Lars Hernquist\altaffilmark{1}, 
Philip F. Hopkins\altaffilmark{1},
Brant Robertson\altaffilmark{1},
and Volker Springel\altaffilmark{3}}

\altaffiltext{1}{Harvard-Smithsonian Center for Astrophysics,
60 Garden Street, Cambridge, MA 02138, USA}
\altaffiltext{2}{Carnegie-Mellon University, Dept. of Physics,
5000 Forbes Ave., Pittsburgh, PA 15213, USA}
\altaffiltext{3}{Max-Planck-Institut f\"{u}r Astrophysik,
Karl-Schwarzchild-Stra\ss e 1, 85740 Garching bei M\"{u}nchen, Germany}

\begin{abstract}

We examine X-ray emission produced from hot gas during collisions and
mergers of disk galaxies.  To study this process, we employ
simulations that incorporate cosmologically motivated disk-galaxy
models and include the effects of radiative cooling, star formation,
supernova feedback, and accreting supermassive black holes.  We find
that during a merger, the colliding gas in the disks is shock-heated
to X-ray-emitting temperatures.  The X-ray luminosity is spatially
extended, rises during the initial stages of the merger, and peaks
when the galactic centers coalesce.  When a physical model for
accreting black holes is included, the resulting feedback can drive
powerful winds that contribute significantly to the amount and
metallicity of hot gas, both of which increase the X-ray luminosity.
In terms of their stellar kinematics and structural properties, the
merger remnants in our simulations resemble elliptical galaxies.  We
find that the X-ray luminosities of the remnants with $B$-band
luminosities in the range $L_B \sim 10^{10} -10^{11}$~\lsun\ are
consistent with observations, while remnants with smaller or larger
masses are underluminous in X-rays.  Moreover, because the majority of
the merger remnants are broadly consistent with the observed scaling
relations between temperature, $B$-band luminosity and X-ray luminosity
we conclude that major mergers are a viable mechanism for producing
the X-ray halos of large, luminous elliptical galaxies.
\end{abstract}


\section{Introduction}
\label{sec:intro}

It is now well established that elliptical galaxies contain
substantial quantities of hot gas \citep[see][and references therein]
{MB03}. Detailed studies of the X-ray emission produced by this hot
gas have revealed that its luminosity is correlated with the gas
temperature, B-band luminosity, and stellar velocity dispersion of the
host elliptical galaxy \citep*{OFP01,OPC03}.  These correlations argue
for an evolutionary link between the process that formed the
elliptical galaxy and the surrounding, X-ray emitting gas.

According to the ``merger hypothesis,'' elliptical galaxies are formed
when two spiral galaxies interact and merge \citep{TT72,T77}.  Thus,
it is possible that disk-galaxy interactions transform spirals
galaxies into ellipticals {\it and} generate their hot gaseous halos.
Using the X-ray satellite $ROSAT$, \citet{RP98} investigated the X-ray
luminosity evolution along the ``Toomre sequence'' \citep{T77}, a
series of eight local galaxies representing different stages of merging,
in chronological order.  Their study found that extended X-ray
emission is produced during the encounter and persists as the nuclei
merge.  Similar results obtain from observations of merging and
interacting systems using the higher resolution $Chandra X-ray
Observatory$ \citep{Fab01,McD03,Huo04}.

A puzzling aspect of the analysis of \citet{RP98} is that the merger
remnants appear to be underluminous when compared to typical
elliptical galaxies, bringing into question the likelihood that merger
remnants evolve into normal ellipticals.  These authors suggested
several methods by which a merger remnant could acquire an X-ray halo,
late infall of tidal material, reacquisition of gas ejected by
galactic winds during the merger, and ongoing mass-loss by stars born
during a merger-induced starburst.  One clue to the origin of the hot
gas lies in the correlation between the X-ray and $B$-band luminosities.
\citet{OFP01}, also using $ROSAT$, found that $L_X\propto L_B$ for
less massive ellipticals, consistent with their X-ray luminosity being
primarily stellar in origin.  However, the X-ray emission from more
luminous ellipticals varies as $L_X\propto L_B^2$, indicating a
non-stellar formation process.

Another clue to the origin of hot gas in spheroidal systems is its
high metallicity.  \citet{HB05} analyzed a sample of 29 X-ray
luminous elliptical galaxies and found that their two-phase models for
the hot X-ray emitting plasma yielded metallicities that were
typically solar or even higher and generally correlated with the
stellar metallicity.  This result indicates that the gas that makes
up the hot coronae surrounding ellipticals has been recycled from
star-forming gas and likely once resided within the dense central
regions of the galaxies.  This also suggests that a significant amount
of energy was injected into this gas in order to expel it from the
depths of the potential well and redistribute it throughout the
galactic halo.

The work described here is a first attempt to address the viability of
galaxy mergers in generating X-ray emitting gaseous halos consistent
with observed ellipticals.  While numerical simulations have
demonstrated that galaxy interactions can indeed produce remnant
galaxies whose stellar component is reminiscent of local ellipticals
\citep{B92,BH96,HRemI,HRemII,NB03,Rob05fp,SdMH05red,Cox05rot}, 
little is known theoretically
about the evolution of the hot gas component in these events.
\citet{BH96} considered simulations in which the gas was not allowed
to cool radiatively and showed that in this limit it would be mostly
shock heated to the virial temperature of the remnant, leaving a
corona of hot, X-ray emitting material.  Later, \citet{Cox04} showed
that the amount of hot gas is dependent on the interaction orbit, i.e.,
more radial encounters produce more hot gas.  Yet the ability of this
hot gas to produce X-rays, the luminosity of the X-ray emission, and
the correlation to the remnant stellar properties are all open
questions.  In particular, these studies did not attempt to model all
feedback processes in detail.  Recently, \citet*{SdMH05} have
developed methods for modeling feedback from both star formation and
black hole growth in galaxy mergers.  This addition to our galaxy
modeling has significant implications for the X-ray evolution as the
feedback energy from an accreting black hole can generate a
substantial amount of hot gas through the production of a large-scale
galactic wind.

The rest of this paper is organized as follows.  In
\S~\ref{sec:sim} we describe the numerical simulations
upon which are results are based.  In \S~\ref{sec:onemerger}
we detail the X-ray emission produced during and after
the collision of two gas-rich disk galaxies.  We follow this by
showing how the global properties of X-rays and hot gas
depend on progenitor mass and gas fraction in \S~\ref{sec:scalings}.
\S~\ref{sec:disc} discusses our results, specifically
addressing the X-ray luminosities of our merger remnants
and the generation of hot gaseous halos observed in luminous
elliptical galaxies.  Finally, we conclude in \S
\ref{sec:conc}.

\section{Merger Simulations}
\label{sec:sim}

To simulate mergers of disk galaxies, we employ {\small GADGET2}
\citep{SpGad2}, which is based on a ``conservative-entropy''
formulation \citep{SHEnt} of smoothed particle hydrodynamics (SPH)
that conserves both energy and entropy \citep[unlike earlier versions
of SPH; see e.g.,][]{H93sph}, while improving shock-capturing.  The
code includes the effects of radiative cooling and heating by a
uniform UV background, and star formation, supernova feedback, and
metal enrichment are treated in the manner detailed in \citet{SH03}.
Within this formalism, the sub-resolution physics of star formation
and supernova feedback acts to maintain star-forming gas at a
temperature specified by an effective equation of state \citep[see
e.g.,][]{Rob04}.  As an extension of this star formation and feedback
model, \citet{SdMH05} introduced an additional parameter $q_{\rm
EOS}$, so that the effective equation of state can be varied between
that for an isothermal gas at $T_{\rm eff} = 10^4$ K, $q_{\rm EOS}=0$,
and the ``stiff'' Springel-Hernquist equation of state, which has an
effective temperature $T_{\rm eff} \approx 10^5$ K, $q_{\rm EOS}=1$.
In the work presented here, $q_{\rm EOS}$ was set to 0.25 resulting in
a mass-weighted temperature of star-forming gas $\sim10^{4.5}$~K.

Our simulations also allow the presence of a supermassive black hole
(BH).  BHs are represented by ``sink'' particles that can accrete
neighboring gas at a rate given by the Bondi-Hoyle-Lyttleton
approximation with an imposed upper limit equal to the Eddington rate.
A small fraction (typically 5\%) of the bolometric luminosity
(0.1$\dot{M}c^2$, assuming an accretion efficiency of 10\%) thermally
couples to the gas surrounding the BH.  These parameters are selected
so that simulations of disk-galaxy mergers successfully reproduce the
normalization of the $M_{BH}$-$\sigma$ relation \citep*{dMSH05}.  The
BH, when included, has a seed mass of $\approx 10^5$~\msun, although
we find that the outcome is insensitive to this choice.  A more
complete discussion of the methodology is provided in \citet{SdMH05}.

The mergers used throughout this paper are identical to those 
in \citet{Cox05rot} and are described in more detail there and
in \citet{Rob05fp}.
A full description of the construction of
these galaxies is given in \citet{SdMH05}; here we simply outline the
parameter choices for the simulations discussed in this work.  The
composite system contains an exponential disk embedded in a dark
matter halo.  For simplicity we do not include a spheroidal bulge.
The dark-mater halo is
initialized with a \citet{H90} profile, a concentration $c=9$, and
a spin parameter $\lambda=0.041$.  In Section~\ref{sec:scalings} we
investigate hot gas production for mergers between equal mass galaxies
with a range of masses, but much of our discussion is based on a
fiducial case where the progenitor galaxies have properties
similar to the
Milky Way.  The virial velocity in this case
is $V_{200}=160$~kms$^{-1}$,
resulting in a halo mass of 1.4$\times10^{12}$ \msun.
The exponential disk composes 4.1\% of the total mass, 
with a fixed fraction $f$ in a collisional gaseous component.  Our
fiducial simulation fixes $f=0.4$.
The compound galaxy is realized with 500,000 particles to represent
the dark matter and 50,000 to represent the disk.  Since the disk 
comprises both stellar and gaseous components, a fraction
$f\times50,000$ are consider gas and the remainder represent the 
collisionless stellar disk.

\begin{figure*}
\plotone{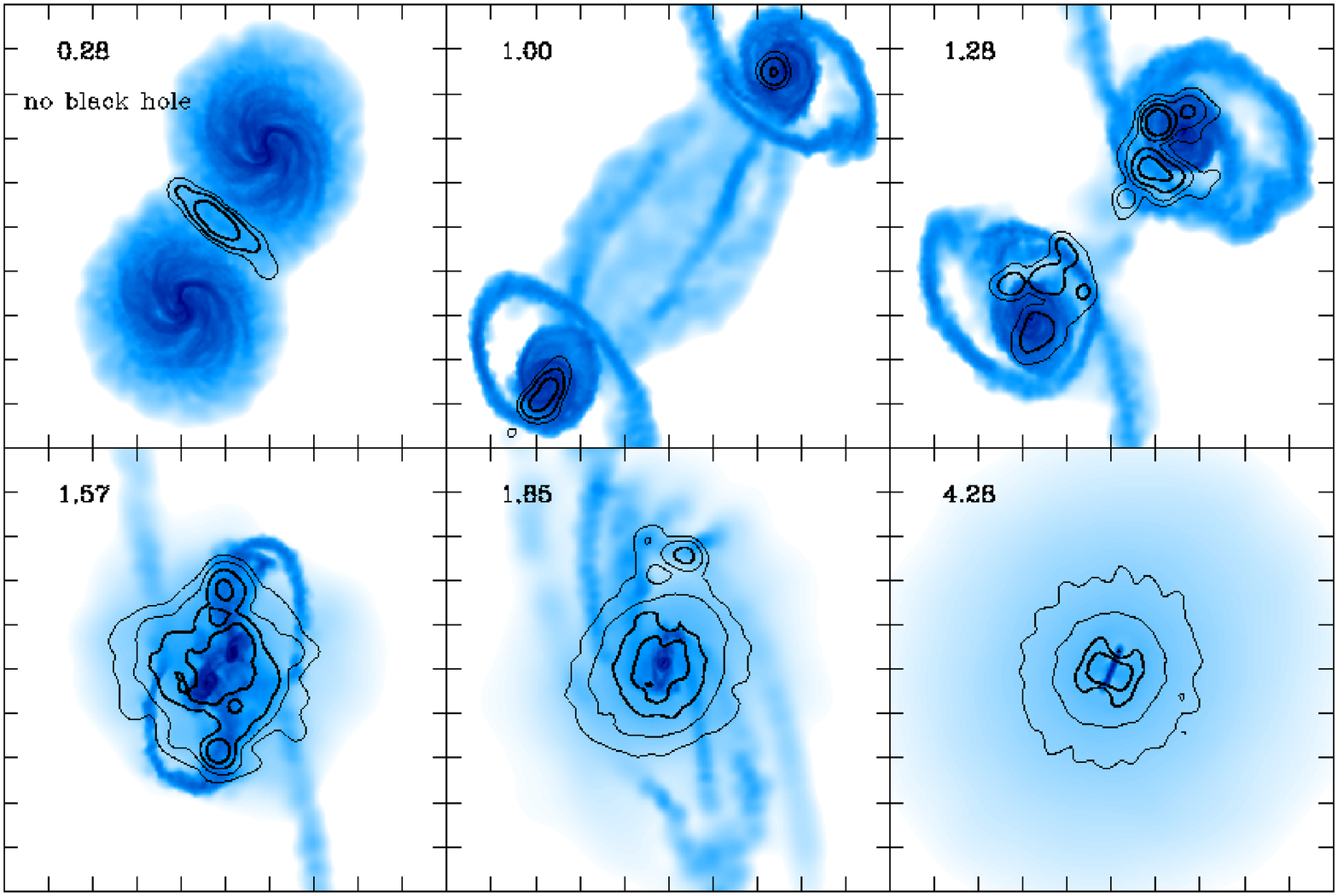}\\
\plotone{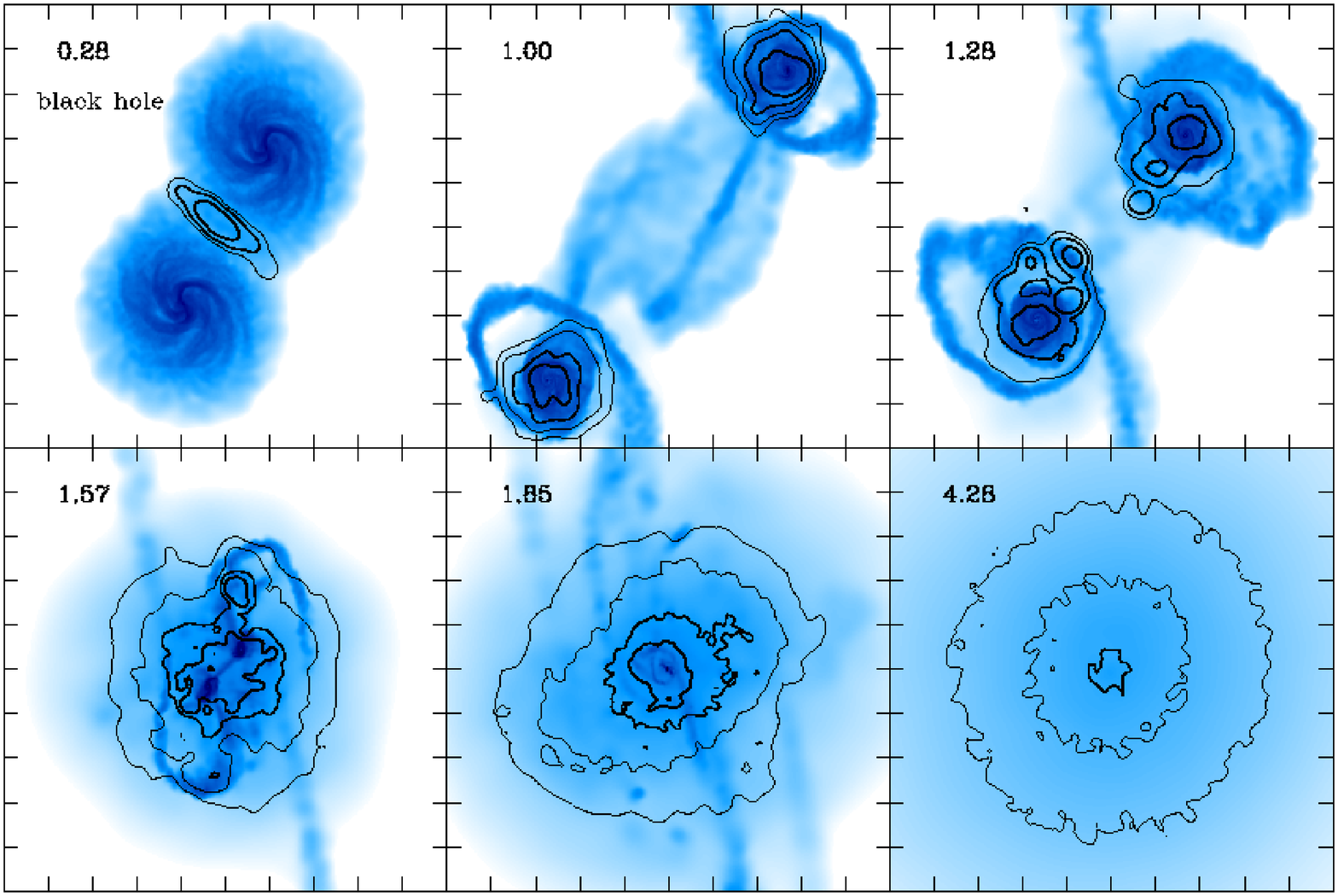}
\caption{
Time sequence of the projected gas density, shown in
blue scale, with X-ray contours overlaid (in black).
The top six panels demonstrate the collision for our
standard, no-black hole merger, while the bottom six
panels are the case in which a
central supermassive black hole is included.
Each panel is 100 kpc square and the color and X-ray
contour scales are identical.
The simulation time,
in Gyr, is indicated in the top left of each plot, and
the merger occurs at $T\approx1.6$~Gyr.
}
\label{fig:images}
\end{figure*}

Once the model galaxies are generated, we collide two disks by placing
them on a parabolic orbit with an initial separation of $143$ kpc and
a pericentric distance of $7.1$~kpc, resulting in a nearly radial
collision.  To investigate the consequences of black hole
growth and feedback, we have performed simulations that include
centrally accreting BHs, and are labeled as such, and others that
do not that are referred to as ``standard''.
Figure~\ref{fig:images} shows six representative images of both types of
galaxy mergers (i.e., with and without BHs).  Details of the morphology,
dynamics and resulting star formation can be found elsewhere
\citep[see e.g.,][]{BH96,MH96,Sp00,SdMH05,Cox05}.

Figure~\ref{fig:images} also demonstrates the additional diffuse
gas present when the merger includes an accreting black hole.
Owing to the energy put into the gas by the growing BHs,
particularly during the final merger, a significant fraction
of the gas participates in a large-scale galactic wind.  This
``blowout'' phase is very efficient and star formation is almost
completely shut off \citep{SdMH05red}, leaving a remnant that quickly
reddens and evolves nearly passively \citep{Hop05Red}.  Moreover, as
discussed further in the following section, the
expulsion of the gas during the blowout greatly enhances the X-ray
luminosity relative to cases in which no BH is present 
because not only is there additional gas throughout the halo, but
this gas is highly metal-enriched.

While a complete survey of the parameter space that describes the
merger of two disk galaxies is beyond the scope of this work, it is
important to understand how the production of hot gas depends on our
initial assumptions.  To this end, we merge disk galaxies with varying
gas fractions and masses and describe the outcomes in
Section~\ref{sec:scalings}.

\section{A Case Study}
\label{sec:onemerger}

In this section we provide a detailed look at the evolution of hot gas
and X-rays during one merger.  While the X-ray luminosity, gas
temperature and mass depend strongly on the initial halo size/mass and
disk gas fraction (discussed in \S~\ref{sec:scalings}),
the production of hot gas during a galaxy collision is a general
feature of all our simulations.

\subsection{X-ray emission}
\label{ssec:xray}

Our analysis of the simulations assumes that X-rays are produced by
the cooling of hot, diffuse gas.  In reality, the observed X-ray
emission has contributions from stellar sources such as supernova
remnants and X-ray binaries, as well as accreting black holes and the
hot phase of the interstellar medium.  Because the composite emission
from these sources is generated mostly in the dense central regions of
the galaxies, this luminosity may be heavily extinguished owing to the
large column density of intervening gas and dust.  In fact, as
described by \citet{Hop05a,Hop05big}, we find that the central
accreting BHs in merger simulations remain obscured for the majority
of their active lifetimes.  To simplify our analysis, we restrict
ourselves to X-ray emission from the diffuse hot gas since it should
be relatively unaffected by obscuration.

We follow \citet{NFW95} and assume the X-ray luminosity
for each SPH particle can be estimated from
\begin{equation}
L_{{\rm X},i} = 1.2 \times 10^{-24}
      \left(\frac{m_{{\rm gas},i}}{\mu m_p}\right)
      \frac{\rho_i}{\mu m_p}
      \left(\frac{kT_i}{\rm keV}\right)^{1/2}  
      {\rm ergs}~{\rm s}^{-1},
      \label{eq:xlum_i}
\end{equation}
where $m_p$ is the proton mass, $\mu$ is the mean molecular
weight (0.6 for a fully ionized primordial plasma), and
$m_{{\rm gas},i}$, $\rho_i$, and $T_i$ are the mass, density
and temperature of the $i^{\rm th}$ gas particle in cgs units,
respectively.  Equation~(\ref{eq:xlum_i}) assumes that the
primary mechanism for X-ray emission is thermal bremsstrahlung,
an assumption consistent with the zero-metallicity cooling
included in the simulation.  However, metal enriched gas with 
a temperature of $\sim10^6$~K cools primarily through
metal-line emission, a mechanism that is more efficient and
would thus produce a much larger X-ray luminosity.  
In this sense, the X-ray emission computed via Equation
(\ref{eq:xlum_i}) is a lower limit.

In light of the increased X-ray luminosity expected from
metal-enriched gas, we supplant equation~(\ref{eq:xlum_i}) with
X-ray emission calculated
using a \citet[][hereafter RS77]{RS77} code that {\it does} include
metal-line emission.  The gas metallicity is provided by 
the star-formation model \citep[see][]{SH03} assuming a
metal yield of 2\%, and we include emission from 0.1 to 2~keV.

The bolometric luminosity is then calculated according to
\begin{equation}
L_{\rm X,bolo} = \sum_{i=1}^{N_{\rm hot gas}}
      L_{{\rm X},i},
      \label{eq:bolox}
\end{equation}
where the sum is carried out over all gas particles with
density below the critical density for star formation
$\rho_{\rm thresh}\approx10^{-2}$ \msun pc$^{-3}$ and a
temperature greater than $10^{5.2}$~K.  Gas particles that
meet these criteria are fully ionized (for zero metallicity)
and are deemed ``hot'' or ``X-ray'' gas from hereafter.
We note that our low temperature ($10^{5.2}$~K) threshold
for gas to be included in the X-ray luminosity probes
temperatures below current observational limits.  However,
this cut-off does serve as an accurate gauge of the amount
of hot, ionized gas.

\subsection{Time evolution}
\label{ssec:time}

\begin{figure}
\plotone{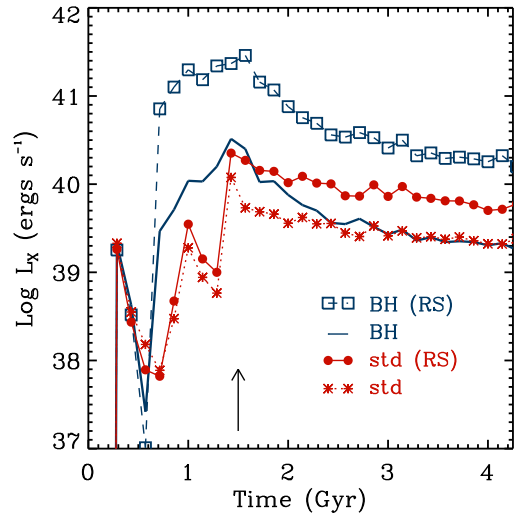}
\caption{
Bolometric X-ray luminosity owing to diffuse hot gas
during simulated major mergers.  Shown are mergers with
(``$BH$'', $blue$) and without (``$std$'', $red$) accreting black
holes.  Each simulation has two curves, one, in which the gas
is assumed to have zero metallicity ($blue, solid~line$, and
$red, asterisks$), and, the other, in which the emission
is calculated using \citet[][RS77]{RS77} models, which include
metal-line
cooling ($blue, open~squares$, and $red, filled~circles$).
The arrow at $T=1.5$~Gyr shows the 
time of the final merger in both simulations.
}
\label{fig:boloxlum}
\end{figure}

Figure~\ref{fig:boloxlum} shows the bolometric X-ray luminosity as
determined by equation~(\ref{eq:bolox}) for the entire major merger.
For both simulations, two X-ray luminosities are shown.
One, where the X-ray luminosity is calculated via equation~(\ref{eq:xlum_i}),
in essence when the gas is assumed to have zero metallicity, and the
second uses the RS77 code to include metal-line cooling.
As expected, the inclusion of metallicity dependent cooling greatly
enhances the X-ray emission produced during the galaxy merger.
In addition, the spatial distribution of the X-ray emission
is shown with overlaid contours in Figure~\ref{fig:images}.

Initially, the X-ray emission from both galaxies is negligible because
we ignore contributions from the star-forming interstellar medium.
X-ray emission begins when the gas disks first interact, in shocks
that lie directly between the two disks (see $T=0.28$~Gyr
panel in Fig.~\ref{fig:images}).  In general, the majority of the
X-ray emission is produced by shock-heated gas, and thus the X-ray
emission essentially tracks shocks that occur during the galaxy 
merger.

Figure~\ref{fig:images} also shows that when the galaxies separate,
the X-ray emission is concentrated around the two galactic nuclei.
This emission resembles that in observed systems that are in a
similar dynamical state, such as the Mice \citep{Read03}.
Figure~\ref{fig:images} also demonstrates that when BHs are present,
additional hot gas is generated by thermal feedback from
BH accretion beginning at $T=0.7$~Gyr.  This results in an $L_X$
that is over an order of magnitude larger than the no-BH case.

The excess X-ray emission from $T=0.7$~to 1.5 is generated by small-scale
winds in the still-separated disks.  These winds are, in general,
collimated and escape perpendicular to the disk plane, similar to
those shown in \citet[][their Fig. 9]{SdMH05}.  In the simulations presented
here, the winds are generated by the energy input by the accreting
black holes and thus are present only in the BH simulations.
Collimated winds such as these are seen in many starbursting systems
\citep{M05,Ru02, VDB05,Str02}.

One of the most prominent features in Figure~\ref{fig:boloxlum} is the
coincidence of the peak X-ray luminosity with the final merger,
regardless of whether or not BHs are present.  This X-ray emission
results from the abundant shocks that occur as the gas disks attempt
to follow the collisionless stellar nuclei.  Gas is heated to
$10^6-10^8$~K and forced out of the central regions owing to shocks and
pressure forces.  Hence, the X-ray luminosity is spatially extended,
as can be seen in Figure~\ref{fig:images} from $T\sim1.5-1.9$~Gyr.

The maximum X-ray emission ranges from $8\times10^{39}$ to
$2.5\times10^{41}$ ergs s$^{-1}$, depending on the presence of a BH
and if metallicity-dependent cooling is included.  These luminosities
are consistent with observations of the X-ray emission seen in
interacting and merging systems
\citep{Fab01,Xia02,McD03,Huo04,Jen05,Read05,SSN05} as well as that
expected on theoretical grounds \citep{Jog92}.  Because the X-ray
luminosity is a direct consequence of the collision between gas in
each progenitor disk, we expect a range of peak luminosities depending
on the gas content in each galaxy, their (relative and cumulative)
mass and the merger orbit.  The mass and gas fraction dependence of
$L_X$ is investigated further in \S~\ref{sec:scalings}.

After the galactic nuclei coalesce, at $T\sim1.6$~Gyr, the X-ray
luminosity decreases.  When no BH is present, $L_X$ drops slowly as
gas settles into hydrostatic equilibrium, and tidal material slowly
returns to the remnant.  However, when an accreting BH is included,
the drop in $L_X$ is more pronounced owing to the ``blowout'' phase
which expels a significant amount of metal-enriched, X-ray emitting
gas.  The eventual fate of gas caught up in the galactic wind is
dictated by its energy relative to the depth of the galaxy potential
well.  In the simulation we describe in this section, 
a substantial amount of gas
is injected with enough energy to completely unbind it from the
remnant.  However, even if a significant amount of gas is ejected from
the system, the blowout does not leave the remnant completely devoid
of gas.  A substantial amount ($\sim75$\%) of the remaining gas
remains bound to the remnant galaxy, is spread throughout the dark
matter halo, and makes up the X-ray corona.

\begin{figure}
\plotone{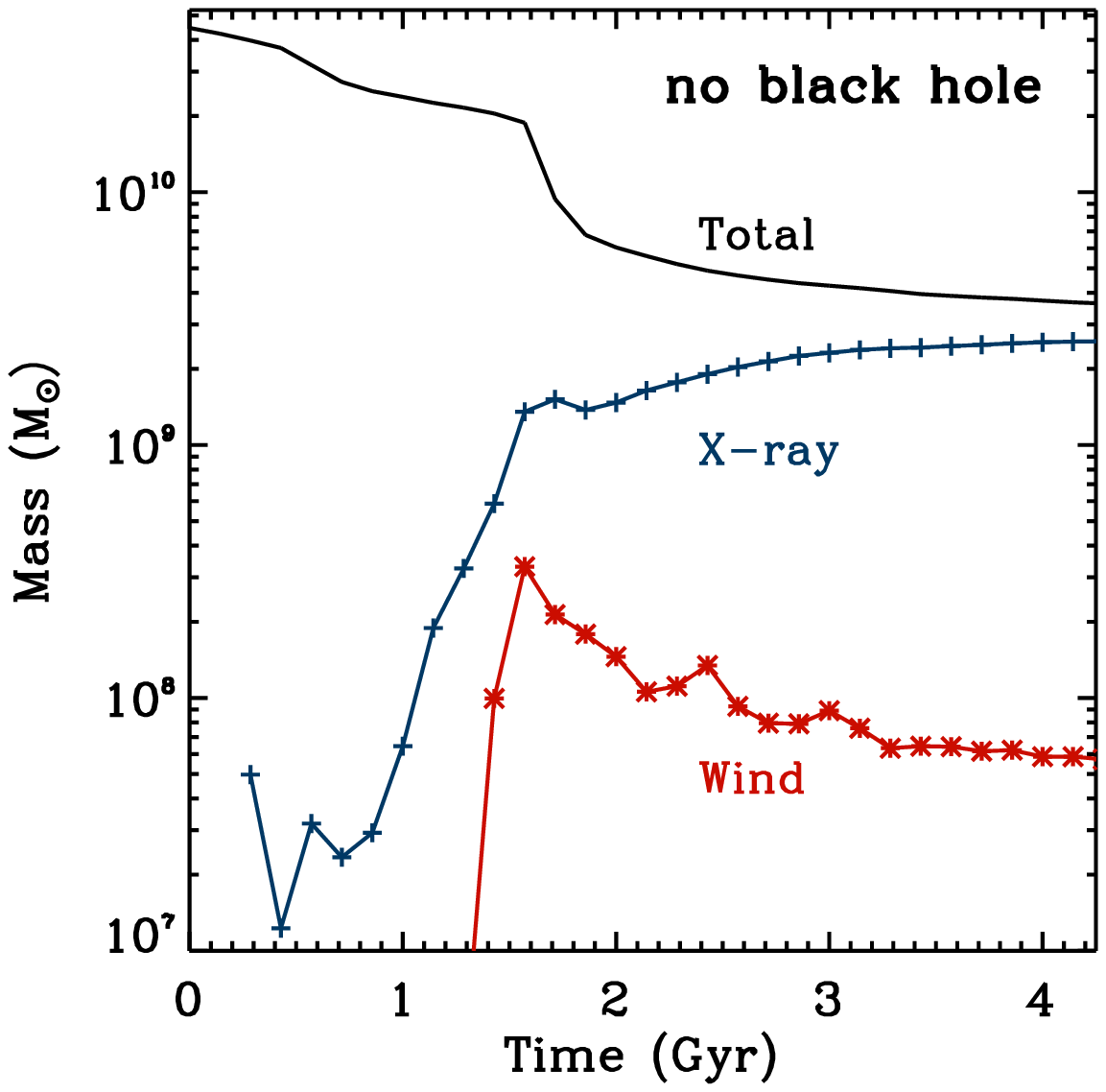}\\
\plotone{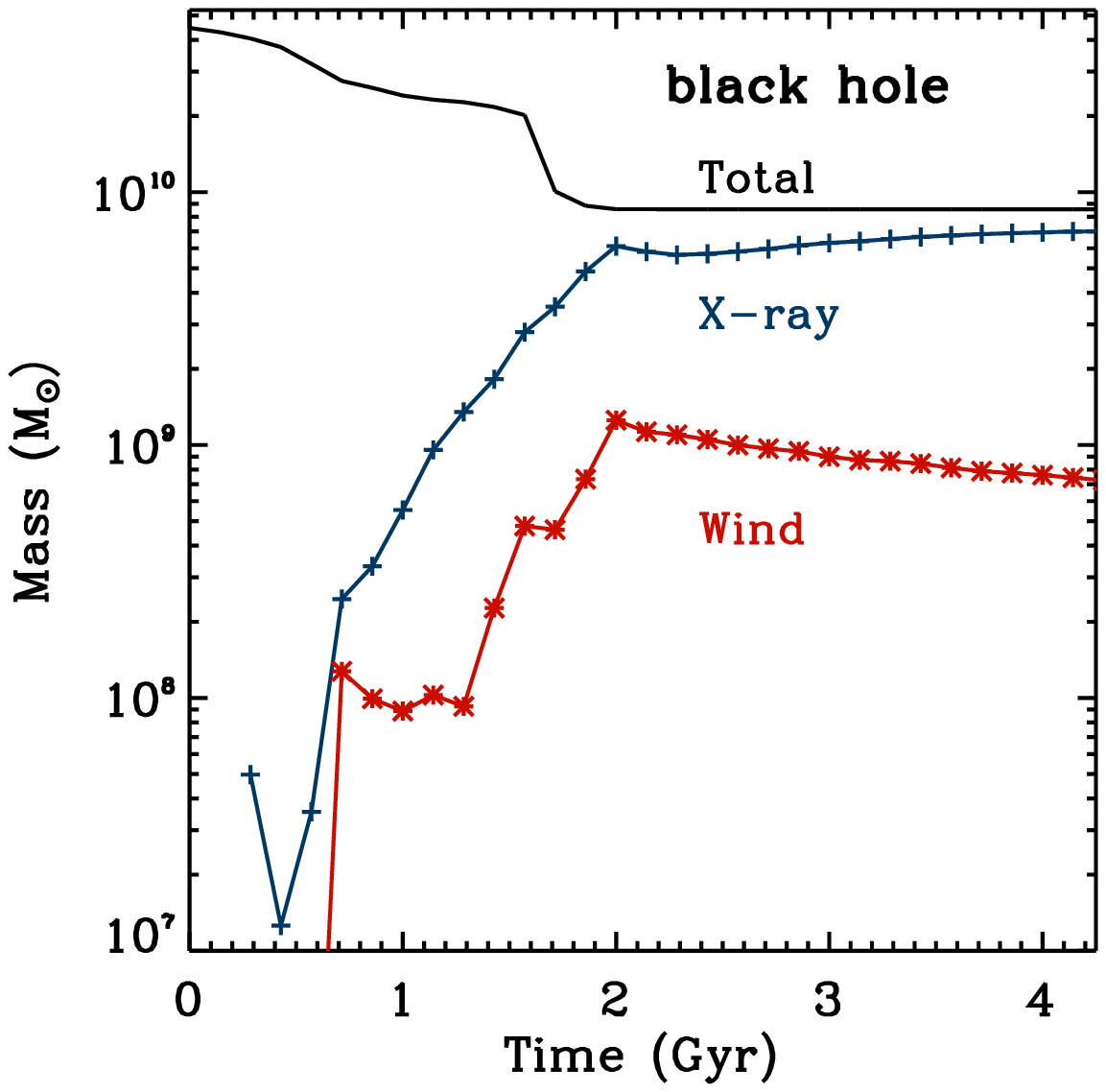}
\caption{
Time evolution of gas mass during the disk
galaxy merger.  Shown with a solid black line is
the total gas mass.  During the simulation, gas is
steadily depleted owing to efficient star formation.
The ($blue$) line with crosses is the mass of gas
which emits X-rays.  The ($red$) line with asterisks
is mass of gas which becomes unbound from the system.
Two plots are shown, one with and the other without a
centrally accreting black hole. 
}
\label{fig:gastime}
\end{figure}

Figure~\ref{fig:gastime} provides an accounting of the gas mass for
mergers with and without accreting BHs.  This figure also shows
the mass of unbound (i.e., $E>0$) gas and the mass of gas
that contributes to the X-ray luminosity.  Gas that is not unbound
or X-ray emitting is typically cold and, depending on its density,
star forming.  Consistent with the findings of \citet{SdMH05red}
there is little cold gas in the remnant that includes a BH and star
formation is efficiently quenched subsequent to the merger at
$T\sim1.6$~Gyr.

Figure~\ref{fig:gastime} demonstrates that the BH has a significant
impact on the amount of X-ray and wind gas.  Owing to the
merger-induced energy input from the BH, and the significant wind
that ensues, the BH system has over 10 times the mass of unbound
gas as compared to the no BH case.  The BH feedback also contributes
the X-ray gas and results in nearly 3 times the amount of mass as
compared to when no BH is present.  At first the differing amounts of
X-ray gas in the two cases seem at odds with the identical X-ray 
luminosity (for emission by thermal bremsstrahlung) demonstrated in
Figure~\ref{fig:boloxlum}.  However, as shown in 
\S~\ref{ssec:3Dprofs} (see Fig.~\ref{fig:gas3D}) the gas
within the inner 10~kpc of the remnant without a BH is over an order
of magnitude denser than the no BH remnant and thus is more efficient
at producing X-rays.  We note that the thermal structure is quite 
similar between the two remnants, as is shown in 
\S~\ref{ssec:rem}.

The hot gas in the remnant relaxes on a timescale that is short when compared to
that of the stellar component of the halo, as reflected by both the regularity
of the X-ray isophotes (see the bottom-right panel in
Fig.~\ref{fig:images}) and the steady X-ray luminosity after
$T=3.0$~Gyr, a little more than 1.5 Gyr after the merger.  In
contrast, stellar material in tidal tails is accreted by the remnant
to produce
long-lived shells, loops, and other fine structure \citep{HS92} by
``phase-wrapping'' \citep{Quinn84,HQ88,HQ89}.  Similar features are
seen around many ellipticals \citep{MC83,Schweizer80,Sch90}.
Because this debris is on nearly
radial orbits, gas in the tails is stripped out and deposited in the
halo or inner regions of the remnant \citep{HW92,WH93}, and the shells
quickly redden.

\begin{figure}
\plotone{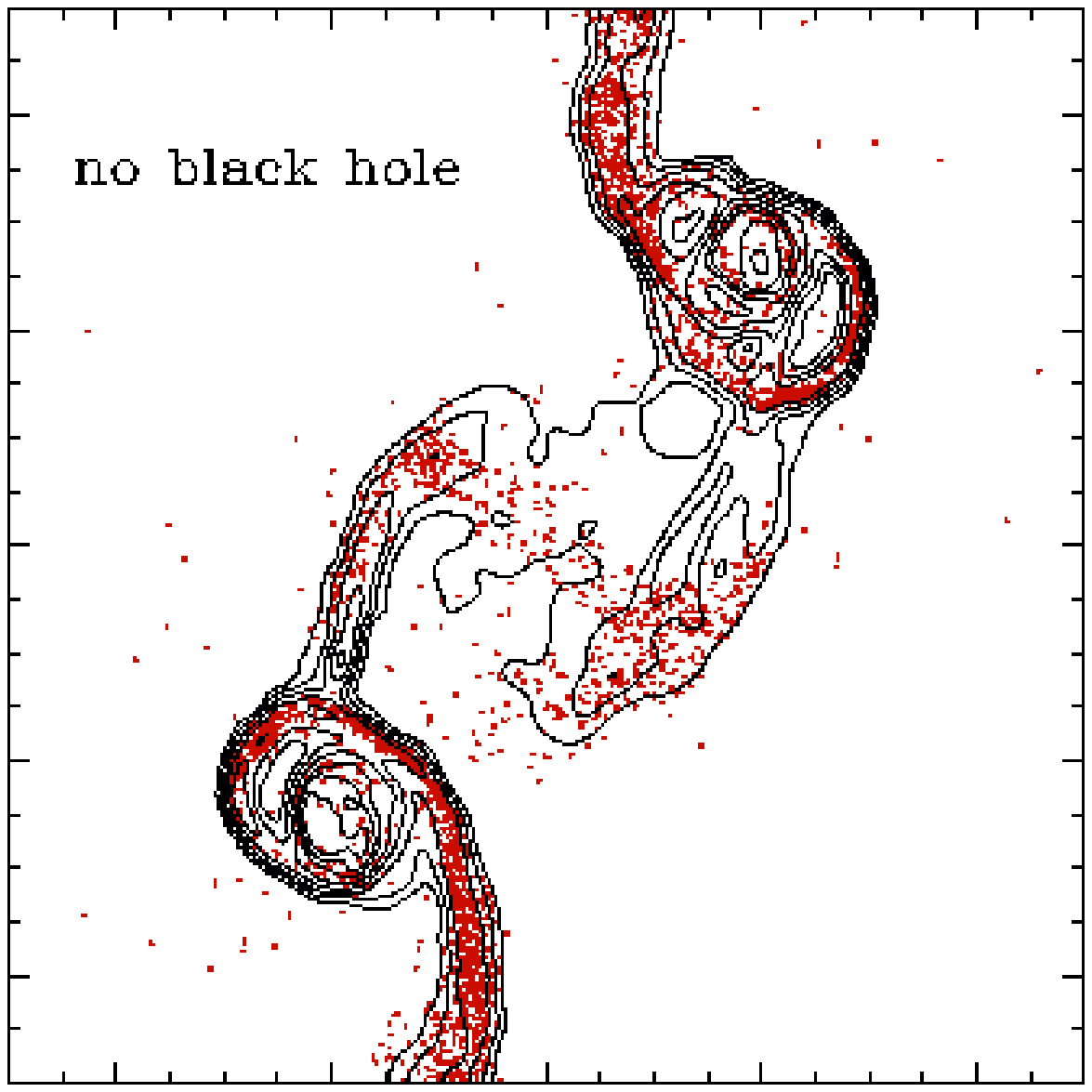}\\
\plotone{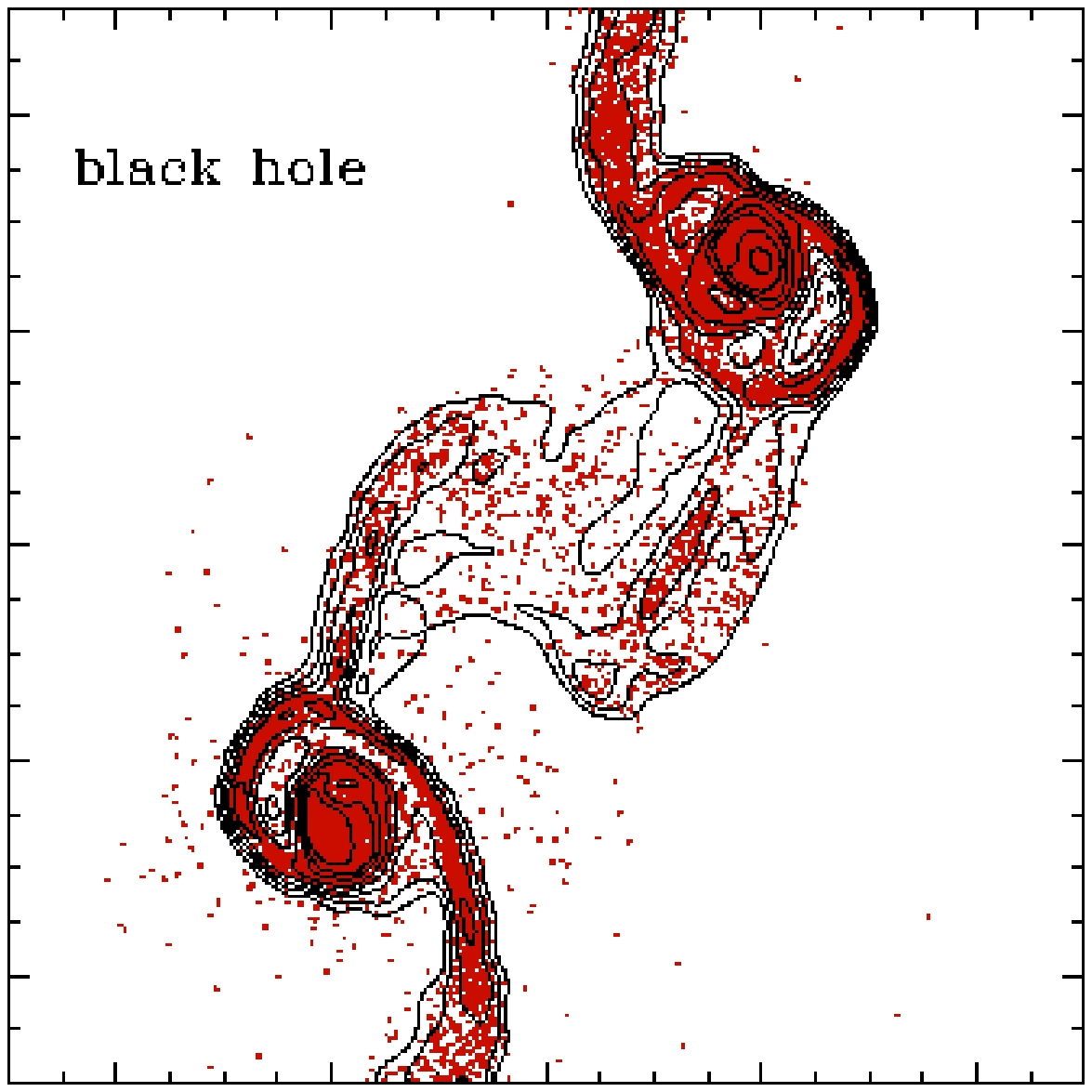}
\caption{
Contours show the gas density subsequent to 
the first passage of the interacting dark galaxies.
Tidal tails and a bridge are clearly present.  Overlaid
are ($red$) dots that denote gas particles that eventually end
up as part of the hot, X-ray emitting gaseous halo.
Two images are shown, one with and the other without a
centrally accreting black hole.
}
\label{fig:xraygas}
\end{figure}

The BH-driven wind is also efficient at dispersing metals.  Because
the blowout is triggered by the centrally located BH, gas that is
ejected from the system has been significantly enriched by metals.  To
demonstrate this more explicitly we refer to Figure~\ref{fig:xraygas}.
Plotted are contours of the gas density at $T\approx0.9$~Gyr, shortly
after the first passage of the two disks.  Overlaid are ($red$)
dots that denote the gas particles that will eventually constitute
the hot X-ray emitting gaseous halo.  When no BH is present, the gas
in the X-ray halo was once almost exclusively composed of loosely
bound tidal material.  In these low-density regions, little star
formation occurs and thus metal enrichment is only modest.  While
this same loosely bound tidal material will make it into the X-ray
halo of the BH remnant, the majority of the X-ray gas in this case
comes from gas that was once more centrally concentrated.  This gas is
spread throughout the disk and has been significantly enriched by star
formation.

For the BH merger shown in Figure~\ref{fig:xraygas}, 
X-ray gas in the remnant has an X-ray
weighted mean metallicity of 0.9 solar.  The unbound wind material has an
even higher metallicity, nearly twice solar.  In this sense, the wind
could be considered ``mass loaded.''  The high metallicity of the gas
suggests that the wind is made up primarily of the densest, most
metal-rich gas. In contrast, without the energy input from a BH, the
X-ray gas is 1/10~\zsun, and the small amount of unbound gas is
less than 1/20~\zsun.  Without the energy input from the BH
essentially the entire metal budget is locked in stars or star-forming
gas.

The dispersal of metals by the wind also has a profound influence on
the X-ray emission.  Because the metallicity of the X-ray-emitting gas is
nearly 10 times larger in the BH remnant as compared to the standard
remnant, the X-ray luminosity is greatly increased.  We show the X-ray
luminosity when metal-line cooling is considered in Figure~\ref{fig:boloxlum}.
We hereafter restrict our analysis to X-ray emission
calculated using the metal-line cooling of RS77.

\subsection{Time evolution of \lx/\lb}
\label{ssec:lxlbtime}

\begin{figure}
\plotone{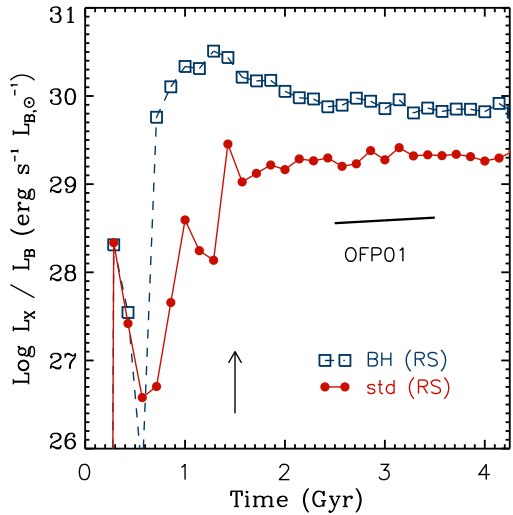}
\caption{
As in Fig.~\ref{fig:boloxlum},
except the X-ray luminosity has been divided by
the $B$-band luminosity.  For clarity, only the
metal-line X-ray emission is plotted. 
Again, the arrow at $T=1.5$~Gyr shows the 
time of the final merger in both simulations,
and the slope shows the observed slope
\citep[OFP01]{OFP01rems}.
}
\label{fig:lxlbtime}
\end{figure}

Because galaxies encompass a range of masses, studies of the X-ray
emission caused by galaxy interactions have typically normalized the
X-ray emission by the $B$-band luminosity of the host galaxy
\citep{RP98,OFP01rems}.  In order to facilitate a more direct
comparison to these studies, we show \lx/\lb\ for our mergers in
Figure~\ref{fig:lxlbtime}.  For simplicity, we ignore the effects of
extinction when computing \lb.  This approximation is likely to
be valid during the post-merger phases but may dramatically
overestimate the luminosity during the highly obscured merger stage.
In any case, many of the features in \lx/\lb\ are very similar to
those of just \lx (shown in Fig.~\ref{fig:boloxlum}).  This result is
not surprising given the moderate gas fraction of our progenitor
disks, and hence the modest fraction of stellar mass that is produced
during the merger.  For the most part, \lb\ slowly decreases over time
owing to an aging stellar population.

Both \lx\ and \lx/\lb\ (Figs.~\ref{fig:boloxlum} and
\ref{fig:lxlbtime}, respectively) demonstrate a characteristic rise
and fall of X-ray emission.  The peak of this emission is coincident
with the merger, a trend that is also seen observationally
\citep[][their Fig. 13]{RP98}.  One discrepancy may be the decline from the
peak which is larger in the simulations than the observations.
However, Figure~\ref{fig:boloxlum} shows a wide range of peak
luminosities depending on the presence of a BH and the metallicity of
the gas.  The luminosities will also fluctuate owing to the gas
content of the progenitor disks as well as the merger orbit, 
complicating a quantitative comparison with the observations.

We can also compare the X-ray luminosity from the simulations
subsequent to the merger to the study of disturbed ellipticals by
\citet{OFP01rems}.  These authors determined that \lx/\lb\ steadily
increases with remnant age when age is measured by the fine
structure parameter $\Sigma$ \citep{SS92} and is zero at the last
merger event.  The slope of the best-fit relation by O'Sullivan et al. is shown as a
short line in Figure~\ref{fig:lxlbtime} (``OFP01'').  The shallowness of this
X-ray evolution demonstrates that the increase of \lx/\lb\ over time
is a very subtle effect and that the hot gaseous halos of elliptical
galaxies are grown over long periods of time.

For the simulation that includes BHs, the \lx/\lb\ ratio is nearly
constant after the merger, a by-product of the equivalent fading of
both \lx\ and \lb.  While we followed each remnant for only a short
period of time after the merger and the isolated nature of the
galaxies in our simulations make the long term luminosity evolution
fairly uncertain, it does not appear that our remnant \lx/\lb\
increases at the same rate as observed merger remnants.

\subsection{Remnant emission}
\label{ssec:rem}

As discussed in \S~\ref{sec:intro} the ``merger hypothesis'' 
posits that the interaction and merger of two spiral galaxies
leads to the formation of an elliptical.  In general,
our simulations support this picture as we are left with
a spheroidal stellar component surrounded by a halo of hot,
X-ray-emitting gas, similar to observed elliptical galaxies.
In this section we investigate this in more detail
by directly comparing the X-ray emission of our remnants
to observations of post-merger and relaxed elliptical galaxies.

\begin{figure}
\plotone{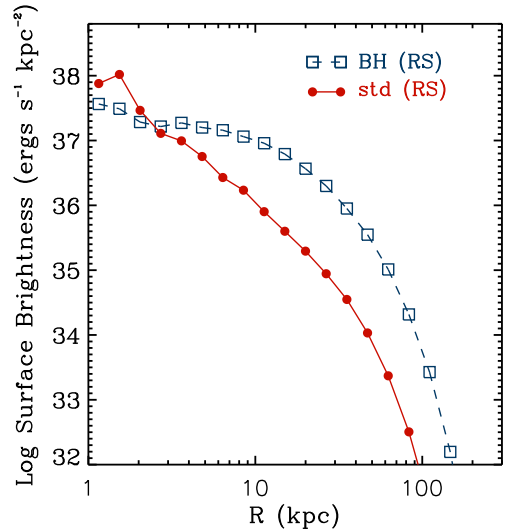}
\caption{
Azimuthally averaged X-ray surface brightness
profile.  The line types are identical to
Fig.~\ref{fig:lxlbtime}.
}
\label{fig:xraysb}
\end{figure}

To begin, we show the X-ray surface brightness in
Figure~\ref{fig:xraysb}.  The emission is calculated using the RS77
metal-line cooling and is azimuthally averaged from 100 random viewing
directions.  The presence of BHs has a significant impact on the shape
and radial extent of the X-ray emission.  Apparently, energy injected
by the BH decreases the emission on small scales ($\leq3$~kpc), and
increases the emission on large scales ($\geq3$~kpc).  In both cases,
however, there is detectable emission to $\sim100$~kpc.

It is common practice to fit the observed X-ray surface 
brightness profiles of elliptical galaxies to a beta model
of the form
\begin{equation}
\Sigma_X(R)=\Sigma_{X,0}
\left[1+(R/R_{\rm core})^2\right]^{-3\beta+0.5},
\label{eq:beta}
\end{equation}
where $\Sigma_{X,0}$ is the central surface brightness, $R_{\rm core}$
is the core radius, and $\beta$ quantifies the slope at large
projected radii.  We fit equation~(\ref{eq:beta}) to the surface
brightness of our remnants and find the best fit to yield
$\beta=1.85$ and $R_{\rm core}=43.3$~kpc for the BH remnant and 
$\beta=2.12$ and $R_{\rm core}=38.6$~kpc for the remnant without a
BH.

We note that the above fit was performed over the entire surface
brightness profile, which includes emission out to $\sim100$~kpc.
At these radii, the X-ray surface brightness is over 5 orders
of magnitude fainter than the central regions.  This large dynamic
range is nearly 3 orders of magnitude more than is probed by 
the standard integration time of observations.  Thus, a more 
realistic comparison can be achieved by restricting our fits to
radii where the X-ray surface brightness is within 2 orders of
magnitude of the peak.  This corresponds to 12 and 50 kpc for the
standard and BH remnant, respectively.  In this case, we find
$\beta=0.80$ and $R_{\rm core}=18.4$~kpc for the BH remnant and
$\beta=0.53$ and $R_{\rm core}=0.73$~kpc for the remnant without a
BH.

The $\beta$-profile fits restricted to the inner surface brightness
profiles are in much better agreement with observations than the
unrestricted fits.  Observed merger remnants have measured $\beta$
values between 0.5 and 1.15 \citep{Nol04,FS95,KF03}.  We also note the much
flatter profile and larger core radius of the BH remnant.  These 
features are observed in several X-ray faint early-type galaxies
\citep{OP04three}.

\begin{figure}
\plotone{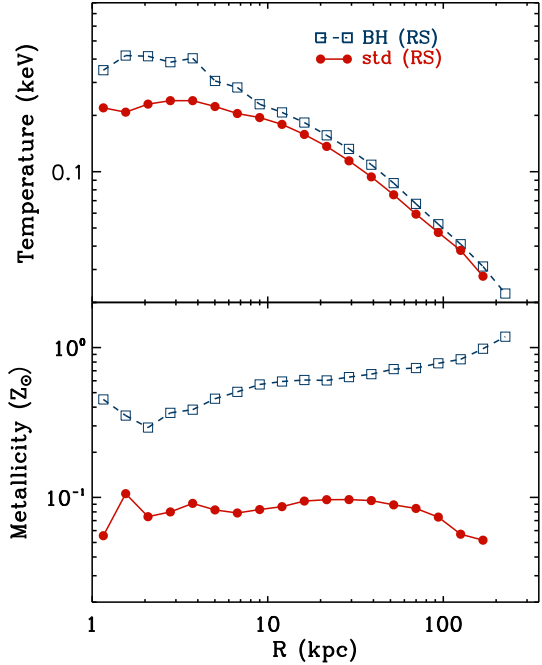}
\caption{ 
Projected temperature and metallicity
profiles for the merger remnants.  Both temperature
and metallicity are X-ray luminosity weighted and
azimuthally averaged.  Line types are as in
Fig.~\ref{fig:lxlbtime}.
}
\label{fig:gasZT}
\end{figure}

In addition to the X-ray surface brightness profile, an increasing
number of galaxies have well determined temperature and metallicity
profiles.  Acquiring this information has been possible with the high
spatial and spectral resolution of $Chandra$ and $XMM-Newton$, as well
as the proper accounting of point sources and complex modeling of the
hot X-ray emitting plasma.  In particular, recent work has shown that
multi-temperature plasma models provide a much better fit to the data and
result in a larger estimate for the hot gas metallicity \citep{BF98}.

Figure~\ref{fig:gasZT} shows the temperature ($top$) and metallicity
($bottom$) profiles for our merger remnants.  Regardless of the
presence of BHs, the temperature is approximately isothermal within
$\sim$10~kpc, and a strongly decreasing function at larger radii.
The additional energy injected by the BH uniformly
heats the gas at all radii without affecting the slope.
However, inside of $\sim2.5$~kpc, the temperature 
difference is much larger owing to different cooling times of the hot
gas.  Without a BH, gas has a shorter
cooling time and a cool(er) core has formed.  In contrast, the
energy input by the BH generates a hot(ter) cusp.  We note
that these curves are expected to have some time variability since
gas has cooling times less than the Hubble time in both instances
(see \S~\ref{ssec:3Dprofs}).

In comparison to the temperature, the differences in
the metallicity profiles are much more striking.  As described earlier,
the BH induced galactic wind is ``mass-loaded'' and transports a significant 
amount of metals to large radii.  This metal-rich wind
results in an {\it increasing} metallicity profile.  The largest
radial bin is supersolar and the most metal rich.  As we showed in 
\S~\ref{ssec:time}, the trend of increasing metallicity with
radius continues to the unbound wind material as well.
Without a BH, metals are retained very near to where they are produced
and hence the metallicity profile is flat or slightly decreases.

The majority of the spatially resolved data exists for X-ray luminous 
early-type galaxies.  These systems typically have nearly isothermal
temperature profiles and metallicity gradients, although there is very
often a slight decrease in temperature at small or large radii, or both
\citep{OP04,OP04three,SSC04,KF03,Mus94}.  At first glance, the profiles
presented in Figure~\ref{fig:gasZT} appear to be at odds with the
observed trends.  However, as we mentioned previously, restricting our
analysis to the inner regions would render the temperature profile much
closer to isothermal.  The metallicity gradient may be indicative of
mixing between metal-rich gas produced during the merger and diffuse
low-metallicity gas that is accreted by the galaxy, a process that is
not included in the present simulations.  We also note that X-ray 
luminous galaxies are
generally large ellipticals, and thus our comparison is best directed
toward merger remnants, which, because they are X-ray faint, do not
usually have measured temperature or metallicity profiles.

One generic feature of observed ellipticals and merger remnants is the
near solar metallicity of X-ray gas \citep{Nol04,KF03,HB05}.  This is
generally reproduced in the simulations that include a BH.
Because the no-BH merger produces
an X-ray halo of low metallicity, this suggests a mechanism to produce
the low metallicities observed in a few X-ray faint
ellipticals \citep{OP04three}.  This could occur if black hole growth is
inefficient, as would be the case if the progenitors are gas-poor, or
if the emission is so faint that one could only detect the central part
of the rising metallicity profile present in the BH remnant.

\subsection{Spherical Profiles}
\label{ssec:3Dprofs}

\begin{figure}
\plotone{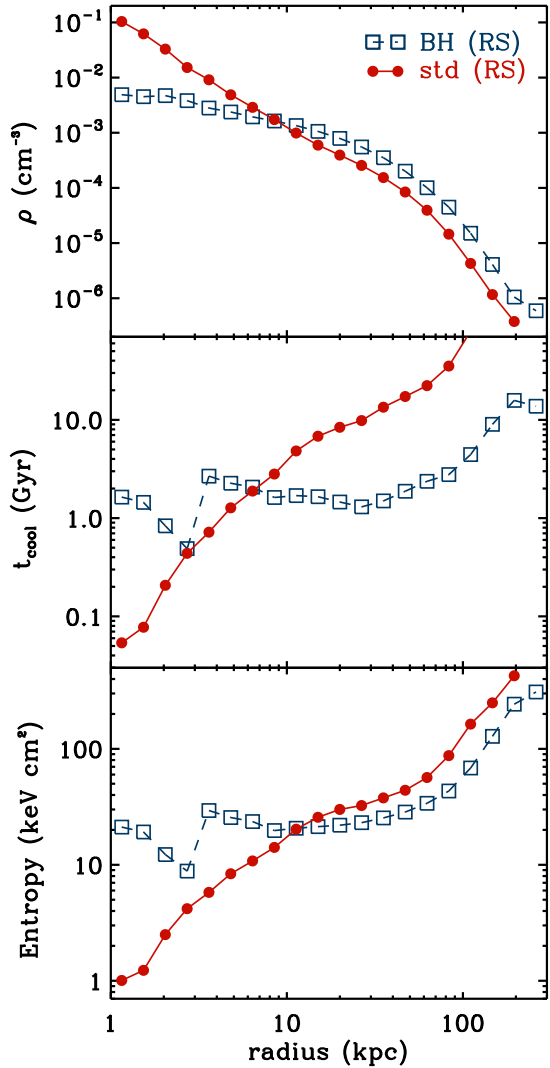}
\caption{
Spherically averaged density, cooling time,
and entropy for X-ray emitting gas.
Line types are as in Fig.~\ref{fig:lxlbtime}.
}
\label{fig:gas3D}
\end{figure}

Figure~\ref{fig:gas3D} shows the spherically averaged
gas density ($top$), cooling time ($middle$), and entropy ($bottom$)
for X-ray gas and
provides a general picture of the state of hot gas in
the merger remnants.  While the remnant profiles are 
quite similar regardless of the 
presence of BHs, there is a distinct
transition at $\sim10$~kpc.  Inside of this radius,
gas in the no-BH remnant has a higher density, and thus 
a significantly shorter cooling time, and lower entropy.
Outside of this radius, the trend reverses, mirroring
what was seen in the X-ray surface brightness profiles
in Figure~\ref{fig:xraysb}.

It is noteworthy that the temperature and metallicity
profiles shown in Figure~\ref{fig:gasZT} do not show
a break.  In this case, the BH remnant has a higher
temperature and metallicity at {\it all} radii.  Together,
these trends suggest that feedback from the BH has pumped
energy and metals into the gaseous component and 
affected diffuse gas at all radii.  In
response to this injection of energy, gas flows outward
(and some even escapes in a wind), filling more of the
dark matter halo than when no BH is present.

Even though gas that is 
located at $>10$~kpc is relatively similar
in temperature and density in both the BH and no-BH cases,
there is a significant difference in the cooling times,
caused by the higher metallicity of gas in the BH remnant.
In fact, much of the existing X-ray halo would cool within
the next several Gyr in this model.

Finally, the spherical gas profiles display
kinks at $\sim3$~kpc.
While very pronounced in the cooling time and entropy
profiles, this kink exists in some form in all the profiles, both
projected and spherical.  Features such as this are quite
common in the simulations with BHs and result from a small
number of residual cold gas clumps within the inner few kpc.

%
\section{Scaling with Progenitor Gas Fraction and Size}
\label{sec:scalings}

Because hot gas is generated by shocks that attend the disk-galaxy
collision, and is bolstered by feedback from black hole accretion,
it is natural to expect that the X-ray emission will depend on the
amount of gas present in the progenitor disk galaxies as well
as their masses.  To understand this relationship better, we 
perform a series of merger simulations that span a range of
progenitor gas fractions and masses as well as the orbits by which
the disk galaxies merge.

All the progenitor disk galaxies used for this parameter exploration
have identical halo properties such as concentration, spin parameter,
and mass fraction.  While holding these properties constant, we 
systematically change the disk gas
fraction or virial velocity, or both.  In total, we construct 
three halos smaller ($V_{200}=56$, 80, and $115$~km~s$^{-1}$) than
our fiducial model ($V_{200}=160$~km~s$^{-1}$) and three that are larger
($V_{200}=225$, 320, and $500$~km~s$^{-1}$).  These virial
velocities correspond to total masses spanning 5.8$\times10^{10}$
- 4.2$\times10^{13}$ \msun, a range of nearly 3
orders of magnitude.  For each galaxy
the exponential disk composes 4.1\% of the total mass, 
with a fixed fraction $f$ in a collisional gaseous component.
The compound galaxies are realized with 500,000 particles to represent
the dark matter and 50,000 to represent the disk.  Since the disk 
comprises both stellar and gaseous components, a fraction
$f\times50,000$ are considered gas and the remainder represent the 
collisionless stellar disk.  In what follows we investigate four
values of $f$, 0.05, 0.2, 0.4, and 0.8.  Black holes are included in
all of the simulations described in this section and the X-ray
emission is calculated using the metallicity-dependent RS77
model.

\begin{figure*}
\plottwo{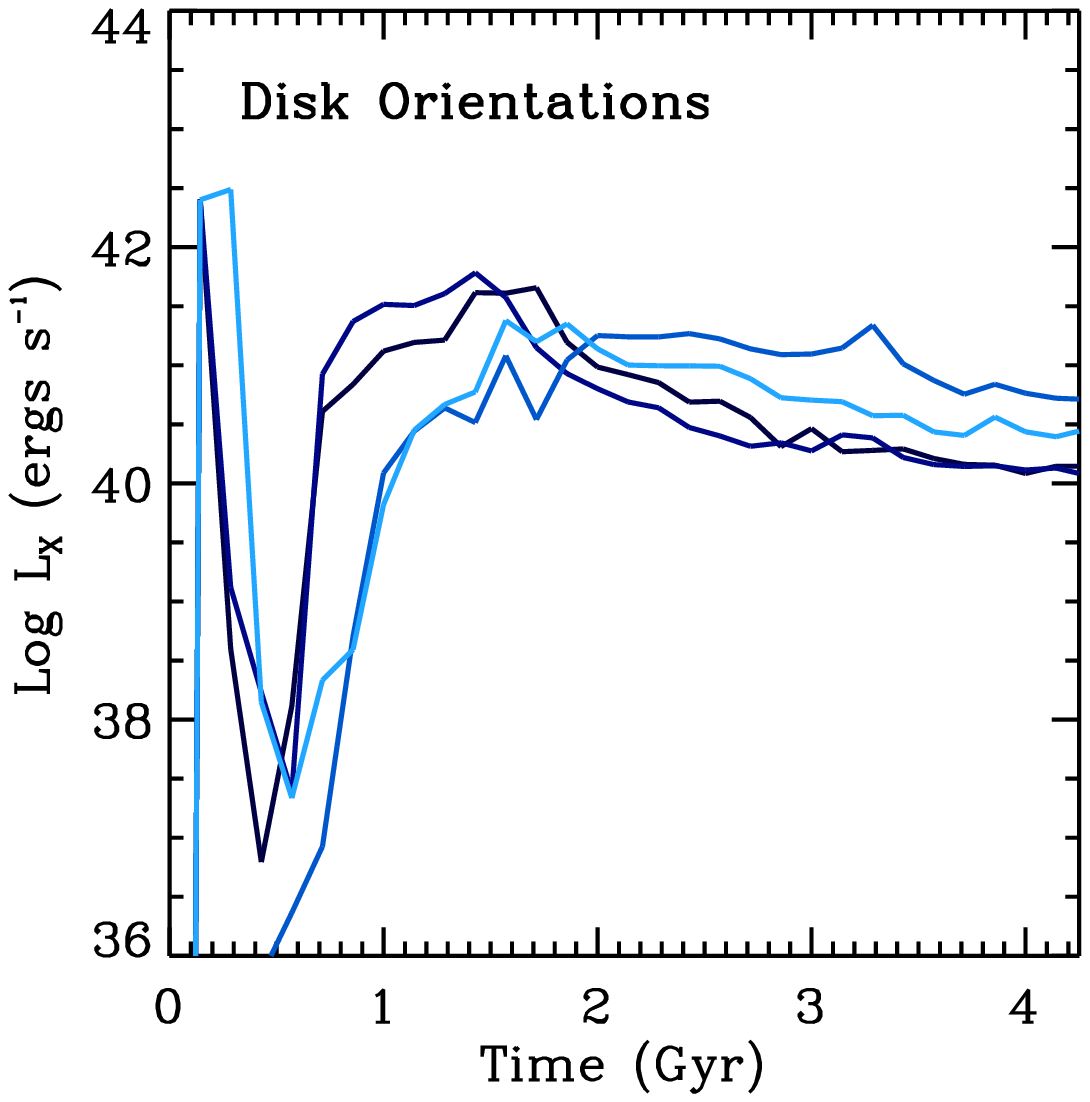}{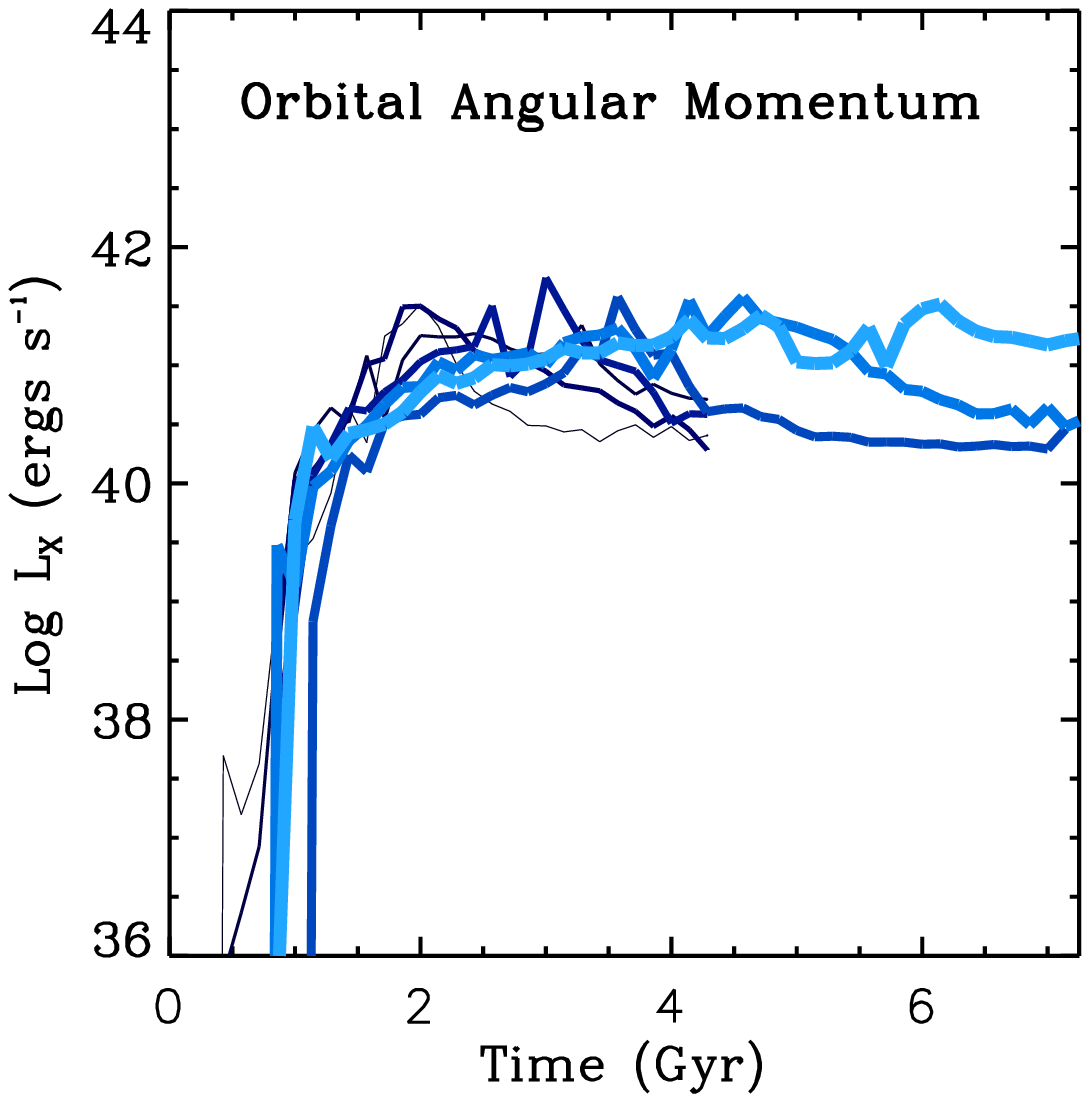}
\caption{
X-ray luminosity calculated using the metallicity-dependent
RS77 model for several disk galaxy major mergers.
In all cases, the progenitor disks are identical, while their
orientation with respect to the orbital plane ($left$), or
their merging orbit ($right$) has been varied.
}
\label{fig:lxoo}
\end{figure*}

Before we begin our exploration of the effects of varying the 
progenitor mass and gas fraction, we wish to determine if the
fully co-planar progenitor disk orientations and radial orbit
of our fiducial disk-galaxy merger
significantly affect the resultant X-ray luminosity,
discussed at length in \S~\ref{sec:onemerger}.  To this end,
we merged the identical progenitor disks on the fiducial orbit with
three alternate disk orientations.  Besides our fiducial co-planar
orientation ($\phi_1 = \phi_2 = \theta_1 = \theta_2= 0$), we 
also simulated a tilted orientation
($\phi_1 = 30, \theta_1 = 60, \phi_2 = -30, \theta_2= 45$), 
another tilted orientation 
($\phi_1 = -109, \theta_1 = -30, \phi_2 = 71, \theta_2= -30$), 
and a polar orientation
($\phi_1 = 60, \theta_1 = 60, \phi_2 = 150, \theta_2= 0$),
where $\phi$ and $\theta$ are the standard spherical coordinates,
and the subscripts 1 and 2 denote the two
progenitor disks.
The X-ray evolution for these runs is presented in
the left panel of Figure~\ref{fig:lxoo}.  In the right panel of
this figure, we show the X-ray evolution for our fiducial 
progenitor disk and fiducial orientation merging on six orbits with
different angular momenta, one with less and five with more.

Together, the orbits and orientations presented in 
Figure~\ref{fig:lxoo} display a range of X-ray luminosities.  At any
particular time, \lx\ may differ by over an order of magnitude.
Much of this variability is attributable to the effect that orbits
and orientations have on the consumption of gas, and thus the effective
gas fraction at the time the galaxies interact.  An additional
factor is the growth of the black hole, as larger black holes will
input more energy and produce additional hot gas.

\begin{figure}
\plotone{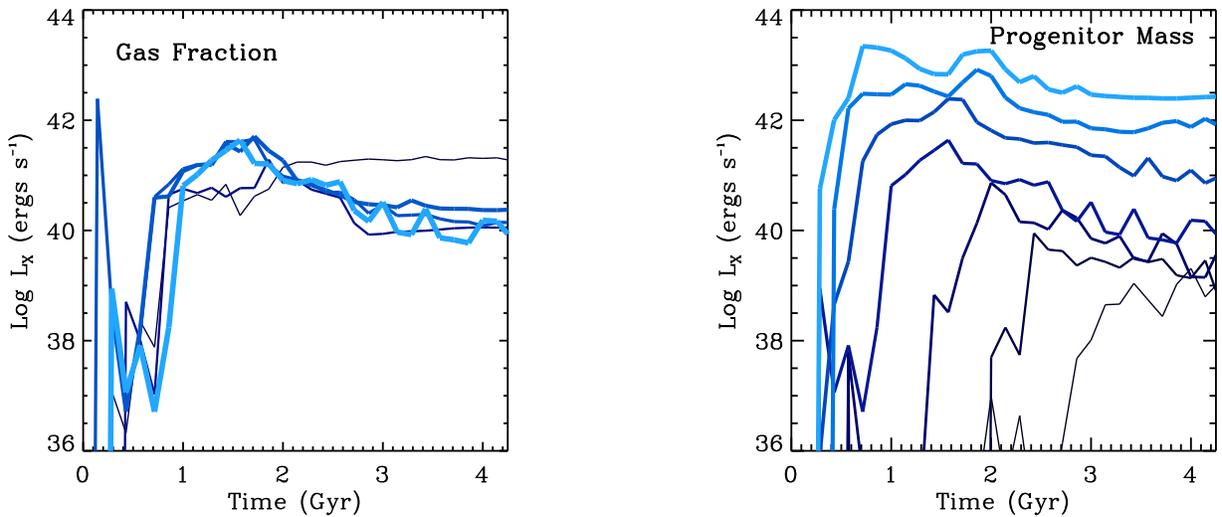}
\caption{
X-ray luminosity (RS77) during a major merger between disks with
varying progenitor gas fractions.  The color and thickness of
each line is proportional to the gas fraction, with lighter,
thicker lines representing the high gas fraction mergers and
dark thin lines the low gas fraction mergers.
}
\label{fig:lxgfs}
\end{figure}

To understand how the progenitor gas fraction influences the
X-ray luminosity without the complications associated with various
orbits and orientations, we resimulate the fiducial merger while
systematically varying the progenitor gas fraction.  The
resultant X-ray evolution is presented in Figure~\ref{fig:lxgfs}.
As the progenitor gas fraction increases, the peak X-ray
luminosity also increases.  There is also an increase
in \lx\ prior to the merger, from $T=1.0$~to 1.8~Gyr, owing to the
larger BHs and additional hot gas produced by their feedback.  Because
the BH feedback drives a powerful wind in the high gas fraction
mergers, the X-ray luminosity drops subsequent to the peak.
There is very little difference in the remnant X-ray emission
for gas fractions greater than 20\%.

For the lowest gas fraction progenitor in Figure~\ref{fig:lxgfs}
the X-ray emission is very small prior to the final merger.
Many of these trends with gas fraction can be
understood in terms of the feedback provided by the BH driving
a large scale wind.  As the gas fraction (and black hole mass)
increase the wind 
becomes dominant and affects a larger amount of gas.  Thus,
the temperature and entropy of the remnant gas steadily increase
in remnants with larger progenitor gas fractions.  These effects
will be the subject of future work that details the winds produced
by black hole feedback (Hopkins et al., in preparation).

\begin{figure}
\plotone{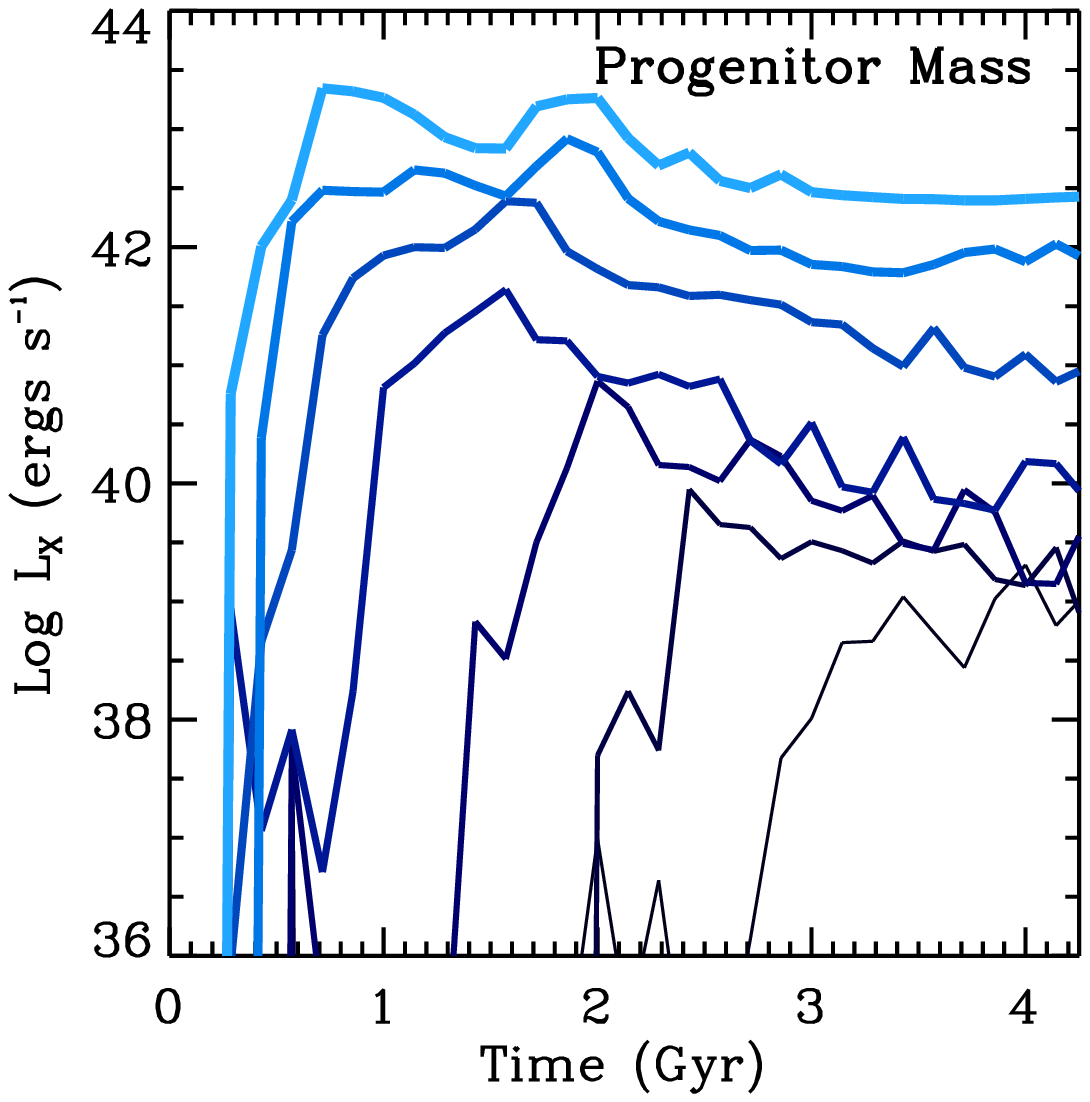}
\caption{
Shown is the X-ray luminosity (RS77) during disk galaxy
major mergers for progenitors of varying mass, from
5.8$\times10^{10}$ \msun to 4.2$\times10^{13}$ \msun.  The
mergers between more massive galaxies produce larger
X-ray luminosities and these curves are drawn with progressively
lighter, and thicker, lines.
}
\label{fig:lxmass}
\end{figure}

The last dependency we explicitly address is progenitor mass.  For
this, we merge galaxies with a variety of masses on orbits that are
self-similar, i.e. the pericentric distance is scaled in proportion
to the size of each disk.  Figure~\ref{fig:lxmass} presents the
X-ray evolution during these mergers.  In all instances, the progenitor
gas fraction was 80\%.

There is a regular progression for more massive progenitors to
produce more X-ray emission.  This is to be expected, if gas settles
into hydrostatic equilibrium in each remnant halo.  There is also
a subtle difference between the maximum X-ray luminosity and the
remnant X-ray luminosity.  Both the most and least massive galaxies
have relatively little difference between the maximum and remnant
\lx, while moderate masses have a large difference.  This trend 
results from two competing effects.  First, the impact of black
hole feedback varies with system mass in a differential manner 
\citep{Hop05slope}, being increasingly more violent for more 
massive galaxies.  Second, the gas consumption is much more
efficient for more massive systems \citep{Rob05fp} and thus 
they have lower effective gas fractions.

\begin{figure*}
\plotone{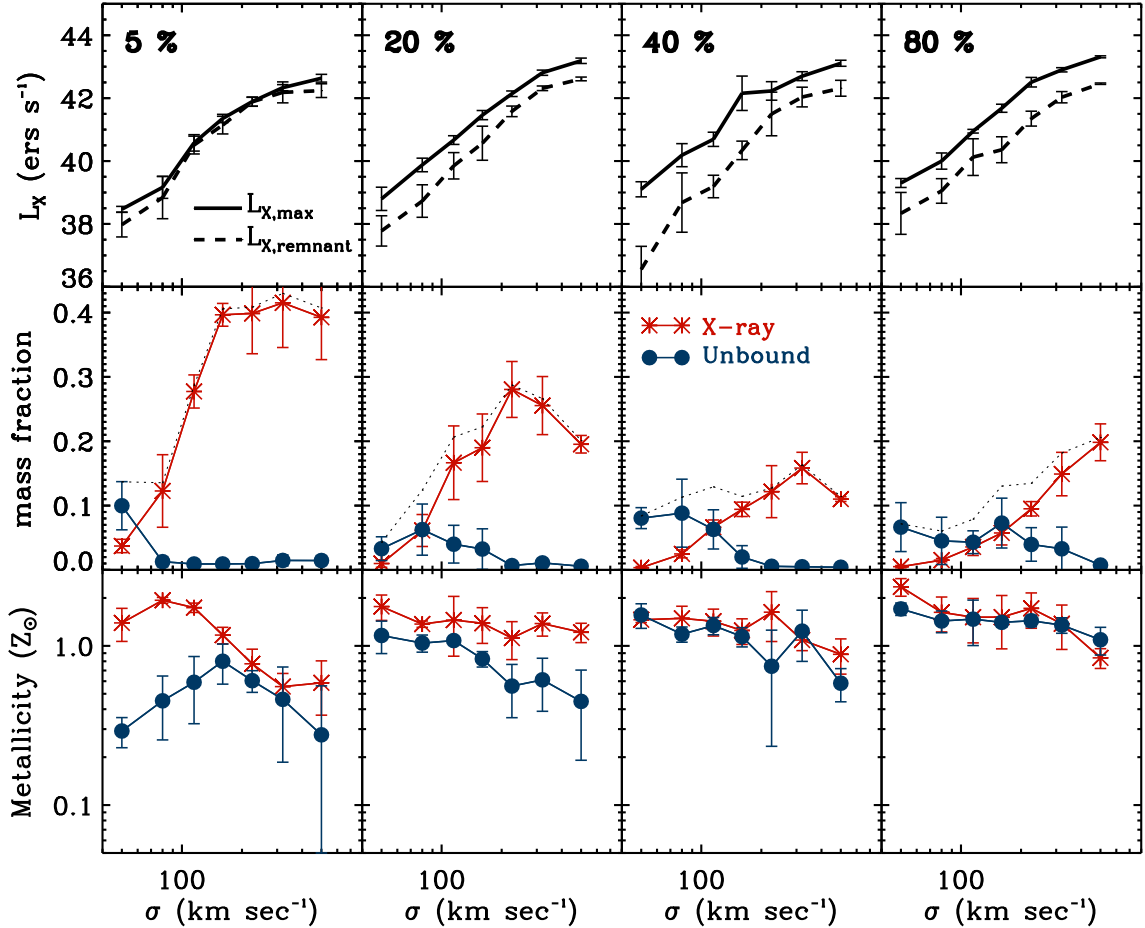}
\caption{
Global properties of the hot gas produced during
disk-galaxy major mergers which include an accreting
BH and its associated feedback.  Each column shows
results for mergers with identical progenitor gas
fractions, indicated in the top panel of each column,
for a variety of progenitor masses, as measured by
the central stellar velocity dispersion of the remnants.
The top row shows the maximum and remnant (RS77) X-ray
luminosity during the merger.  The middle row shows the
fraction of the progenitor gas that is either
physically unbound (i.e., $E>0$) from the remnant or
that constitutes the hot X-ray halo.  The bottom row
shows the average metallicity of the unbound wind
and X-ray gas.
}
\label{fig:scalings}
\end{figure*}

Finally, we summarize the production of X-rays and hot gas during 
the various disk-galaxy major mergers by compiling a number of
global quantities for each merger remnant.  In order to characterize
the X-ray luminosity we determine the peak and remnant \lx\ for each
galaxy merger.  We also find the mass of gas that composes the
hot X-ray halo and the mass of gas that becomes unbound from 
the galaxy potential owing to the black hole feedback-induced
blowout.  These global quantities are
shown in Figure~\ref{fig:scalings} for our entire merger series, which
includes the seven different mass initial conditions, four initial
disk gas fractions, and four disk orientations.  The scatter resulting
from the four initial disk orientations simulated is shown by the 
$1\sigma$ error bars for all quantities.

The top panels of Figure~\ref{fig:scalings} demonstrate what was
also apparent in Figure~\ref{fig:lxmass}, namely, that more massive
galaxy mergers produce more X-ray luminosity.  For a self-similar
model we expect $\lx \sim \sigma^4$, which is close to what is 
present in the simulated remnants.  There is, however, a slight
flattening at high $\sigma$, which reflects the more efficient star
formation and thus less relative gas in these systems.

Apparent in the middle panels of Figure~\ref{fig:scalings} is a trend
for low-mass mergers to produce more unbound material.  This is
opposite to the trend found for X-ray gas mass, where the most massive
systems have a larger fraction.  Both trends hold across all progenitor
gas fractions investigated.  However, there is a systematic increase
in the fraction of unbound gas with progenitor gas fraction.  This
trend reflects the increased wind efficiency induced by the black
hole feedback for these systems.

The bottom panels of Figure~\ref{fig:scalings} show the metallicity of
the X-ray and unbound gas.  In nearly all the remnants, both the unbound and
X-ray gas have solar or supersolar metallicity.  There is a slight
decrease in metallicity at high masses, as these systems have very
short gas consumption timescales \citep{Rob05fp}.  Thus many of the
metals become locked up in the stars, producing a correlation between
remnant mass and stellar metallicity.

\section{Discussion}
\label{sec:disc}

The previous section demonstrates that galaxy mergers produce
significant X-ray emission and that each merger remnant contains hot,
X-ray emitting gas.  However, we have not addressed whether this X-ray
emission is comparable to observed ellipticals.  Observations indicate
that merger remnants are underluminous when compared with elliptical
galaxies and that ellipticals build their X-ray emission slowly, over
the course of several Gyr \citep{OFP01rems}.  In this section we
compare our merger remnants to common scaling relations for elliptical
galaxies and speculate on their future evolution.

\subsection{The \lx-\lb\ relation}
\label{ssec:lxlb}

\begin{figure}
\plotone{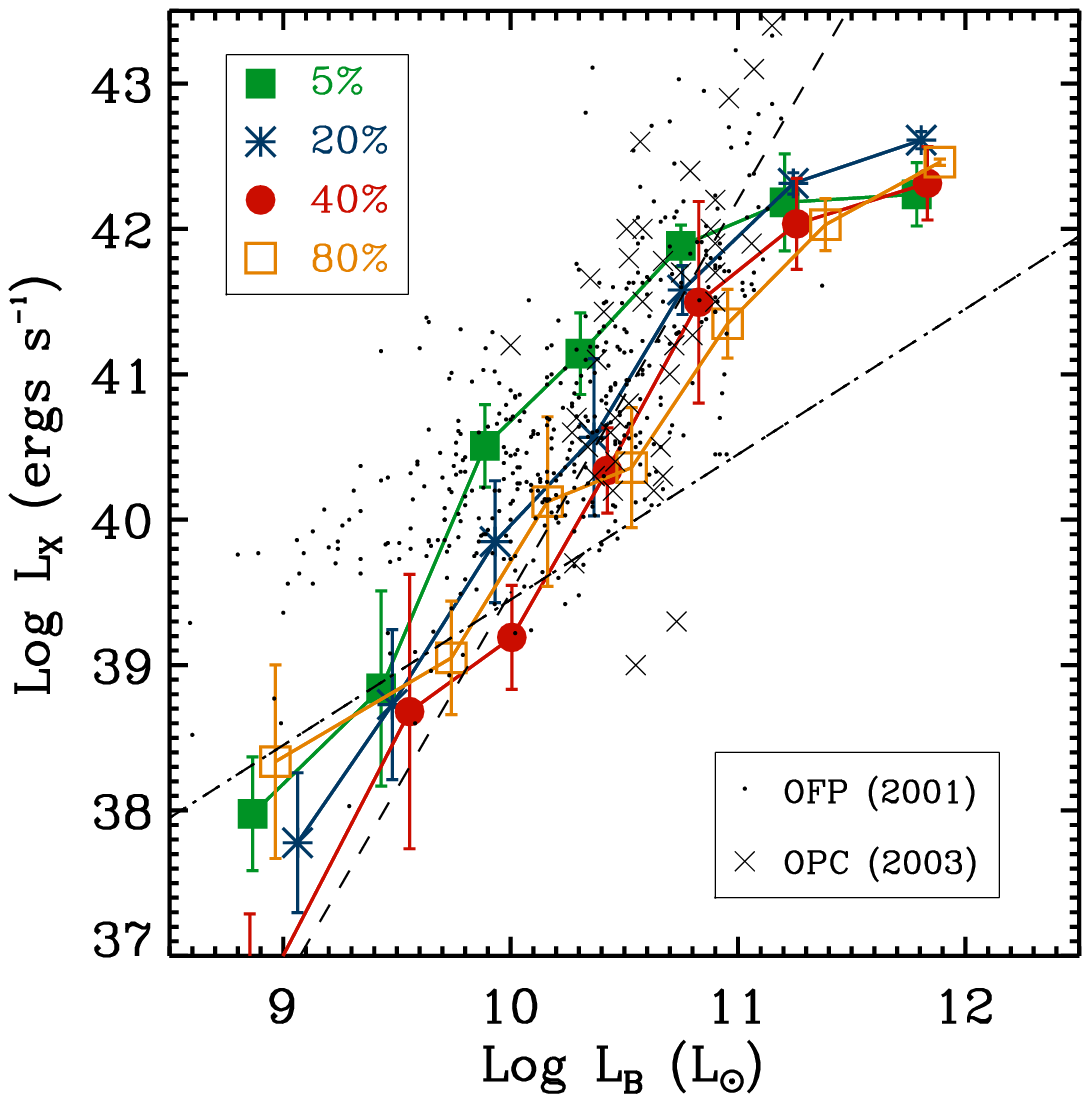}
\caption{
Merger remnants from our entire series of gas fraction and size/mass
simulations plotted in the \lx-\lb\ plane.  These are the same
systems as in Fig.~\ref{fig:scalings}.
The black points
are galaxies from the X-ray catalog of \citet[OFP01]{OFP01}, and
the crosses are X-ray luminous ellipticals from \citet[OPC03]{OPC03}.
The dotted line is the estimated point source contribution
$\log{L_{\rm discrete}}=29.5\log{\lb}$ from \citet{Cio91},
and the dashed line is the
best-fit relation for luminous ellipticals found by OPC03.
}
\label{fig:LxLb}
\end{figure}

One of the most common relations used in the study of X-ray emission
from individual galaxies is the \lx-\lb\ relation.
Figure~\ref{fig:LxLb} shows our entire series of gas fraction and 
size/mass remnants (from Fig.~\ref{fig:scalings}) in the \lx-\lb\ plane.
Also shown on this plot as points are data from \citet{OFP01} and 
luminous ellipticals from \citet{OPC03} as crosses.
The dot-dashed line (with the shallower slope) 
that demarcates the lower bound of
observed galaxies is an estimate of the point source contribution
to the total X-ray luminosity \citep{Cio91}.  
Galaxies that reside near this line have X-ray emission
that is presumably dominated by point sources.  Shown with a 
dashed line of slope $\sim2.7$ is the best-fit relation for 
luminous ellipticals from \citet{OPC03}.

Our merger remnants trace out an arc that passes through the observed
galaxies.  Because we have not included any contribution to the X-ray
luminosity from point sources, the low-mass remnants fall well below
the observations.  This is consistent with the expectation that these
systems radiate little diffuse gas emission or that the diffuse
gas is too cold to be easily detected.  Our moderate \lb\
($\sim10^{10} - 10^{11}$) galaxies are very comparable to observed
ellipticals, while the higher \lb\ systems are underluminous.

The comparison between observed 
ellipticals and our remnants is muddled by two facts, one is the 
uncertain \lb\ owing to
the unknown star formation history of stars that are formed
before we begin our simulation, and the second is the baryon
fraction of our initial disk galaxies, which we assume is less than
one-third of the cosmic mean.

To address the first point, we assume that stars present at the
beginning of the simulation have been steadily formed during the 5~Gyr
prior to our $T=0$.  Metallicities are selected randomly between
solar and 1/10 solar and \lb\ is calculated using the models of
\citet{BC03}.  We also note that the stopping point for our simulations
is arbitrary.  An additional 5 Gyr of evolution would 
decrease \lb\ by another factor of 2.

The second point above, the initial baryon fraction, is particularly
relevant for our most massive systems, whose closest analogs 
are probably groups rather than individual galaxies.  Groups tend to
have baryon fractions closer to the cosmic mean value \citep{San03}
and thus it
may not be surprising that these remnants are underluminous.
We also find it intriguing that X-ray emission is only detected in
groups that have at least one early-type galaxy \citep{Mul03}.
This correlation naturally arises when the X-ray emission and
early-type galaxy are formed in a common process, as we suggest
here.

\subsection{The \lx-\tx~relation}
\label{ssec:lxtx}

\begin{figure}
\plotone{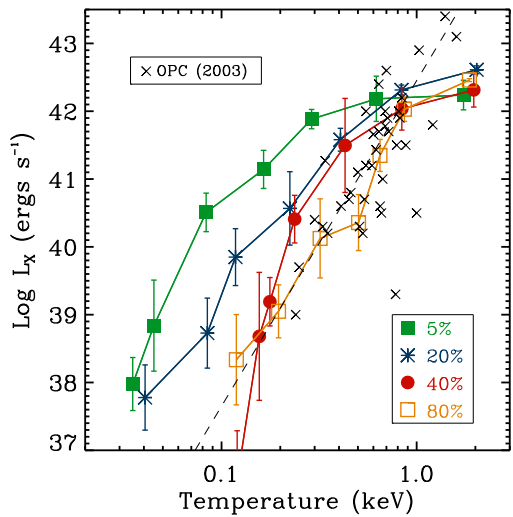}
\caption{
Bolometric X-ray luminosity vs. X-ray-weighted
average temperature for our merger remnants. 
The crosses are X-ray-luminous ellipticals from OPC03,
along with their best-fitting line.
}
\label{fig:LxTx}
\end{figure}

Figure~\ref{fig:LxTx} shows the remnant X-ray luminosity versus
X-ray-weighted average temperature for our merger remnants in
the \lx-\tx\ plane.  The \lx-\tx\ relation is commonly used to 
study X-ray emission from clusters, groups, and more recently,
elliptical galaxies.  The best-fit relation for ellipticals 
appears to have an offset in both slope and normalization from 
the relations found for groups and clusters (which themselves 
are not identical), indicating that these systems have had an 
injection of energy (or entropy) above that expected from purely
self-similar gravitational collapse \citep[see][for a review]{Voit05}.
On this figure we also plot the data of \citet{OPC03},
which consists primarily of luminous ellipticals, as crosses,
and their best-fit relation which has a slope of 4.8.

The majority of the remnants are overluminous when compared to the 
observed \lx-\tx\ relation, while the most massive remnants are
underluminous.  The only remnants that appear to match the observed
data well are the remnants of mergers between disks with 80\% initial
gas fractions.

The discrepancy for the most massive remnants appears to be that
they are simply underluminous, as this is the case for \lx-\tx\ 
as well as \lx-\lb\.  However, most of the other merger remnants
appear consistent with the observed \lx-\lb\ relation, and thus it
is unclear whether they are really overluminous or whether their
temperature is underestimated.  The latter scenario could arise
because of the radial temperature gradient present in most of 
the simulations.  Hence, using a smaller aperture to measure the
X-ray temperature or only detecting the central emission will 
yield higher \tx than that measured here.

\subsection{The $\sigma$-\tx~relation}
\label{ssec:sigtx}

\begin{figure}
\plotone{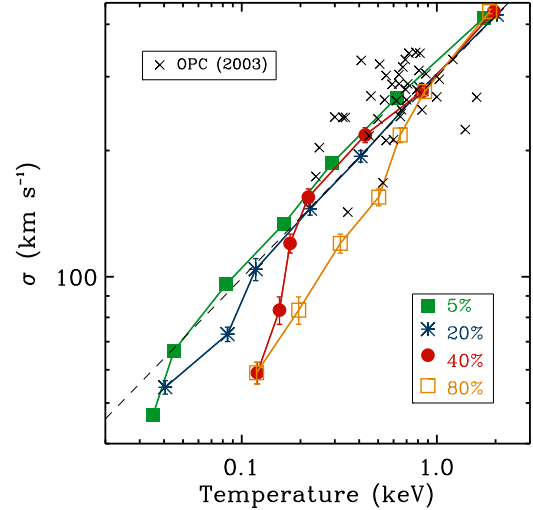}
\caption{
Merger remnants and the central stellar velocity
dispersion $\sigma$ vs. \tx\ plane.
The crosses are X-ray-luminous ellipticals from OPC03.
The dashed line marks equipartition of energy between
X-ray gas and stellar velocity dispersion.
}
\label{fig:SigTx}
\end{figure}

As a final comparison, we plot the X-ray temperature against
the stellar central velocity dispersion in Figure~\ref{fig:SigTx}.
As discussed in \S~\ref{ssec:lxtx}, \tx\ is the mean
X-ray weighted gas temperature and, as shown in 
\S~\ref{ssec:rem}, is a decreasing function of radius.
The central stellar velocity dispersion is measured within the
half-mass radius from 100 random projections.  The 
typical $1\sigma$ error owing to projection effects is 
$\sim$10\%.

Provided that both the gas and stars are relaxed and in 
equilibrium, both $\sigma$ and \tx\ are a measure of the
system potential.  Also shown in Figure~\ref{fig:SigTx} is
a dashed line expected when the gas and stars are 
in equipartition, i.e.,
\begin{equation}
\mu m_{\rm p} \sigma^2 = kT,
\end{equation}
where $\mu$ is the mean mass per particle, and $m_{\rm p}$ is the
proton mass.

Figure~\ref{fig:SigTx} plots $\sigma$ versus \tx\ for our series
of merger remnants.  In general, there exists a tight relation
between $\sigma$ and \tx\ for high mass remnants, signifying
that these systems contain gas that is in hydrostatic equilibrium
throughout the remnant halo.  There exists a trend for low-mass
systems to peel away from the $\sigma$-\tx\ relation.  In
particular, the remnants produced by gas-rich mergers show the 
largest discrepancy.  These remnants are also strongly affected
by the BH feedback (see, e.g., the amount of unbound gas mass in 
Fig.~\ref{fig:scalings}) and their high temperature reflects this
additional energy input.

\subsection{Building an X-ray halo}
\label{ssec:build}

Overall, our model for the production of X-ray emission during
the merger of two equal mass disk galaxies successfully matches the
rise and fall of X-ray emission observed for the Toomre sequence
\citep{RP98}.  In addition, we find that the X-ray luminosity of the
merger remnants span a wide range that depends on the initial disk
gas fraction and size/mass.  Compared to observed early-type
galaxies of equivalent \lb, our models are broadly consistent with the
data for Log(\lb)$<11$~\lsun, and underluminous for more massive remnants.
However, because the hot gas is generally consistent 
with the \lx-\tx\ and $\sigma$-\tx\
relations, we argue that a disk-galaxy merger that includes the feedback
from accreting supermassive BHs is a sufficient mechanism
to produce hot gas in the correct ``state,'' i.e., metal-rich and 
in hydrostatic
equilibrium, but in some cases there is simply not {\it enough} of this gas.

As stated in the introduction, three mechanisms have been proposed
to explain the growth of a luminous X-ray halo for elliptical
galaxies, infalling tidal material, the reacquisition of material
ejected in a galactic wind during the formation of the elliptical,
and stellar mass loss.  We discuss each of these in turn.

\subsubsection{Infalling tidal material}

One definitive conclusion from our simulations is that infalling
tidal material contributes very little to the remnant X-ray emission.
While tidal material continues to rain down on the remnant long
($\sim1$~Gyr) after the nuclei have coalesced, the mass of these 
cold tidal tails is small ($\sim20\%$) compared to the mass of hot
diffuse gas already in place.  Not only is the mass in the tidal
material not sufficient to significantly alter the X-ray luminosity,
but the efficiency at which this infalling cold gas is
converted to X-ray-emitting hot gas is less than $30\%$.  Thus, the
diffuse hot gas is largely in place soon after the merger, as 
demonstrated in the $T=1.85$~Gyr
panel of Figure \ref{fig:images}, and the
tidal material serves only to ``top off'' the amount of hot gas.
This is consistent with the theory that shells around ellipticals
form from accreted tidal debris \citep{HS92}, because the stellar
luminosity of these features is small compared with that of the host
galaxies.

\subsubsection{Re-acquisition of ejected hot gas}

Within the context of the simulations we present, only the mergers
that include BHs and produce a wind have a reservoir of ejected gas
to reacquire.  This is evidenced by Figures~\ref{fig:gastime} and
\ref{fig:scalings}, which show that the galactic wind is ubiquitous in
small-mass halos and depends on the initial disk gas fraction.  In
some cases, the wind contains more mass than that which remains in the X-ray
halo.

Of course, all of our remnants will also accrete pristine gas
through the hierarchical growth of structure, which we do not follow here.
This gas is likely to be acquired by infall from the ambient 
intergalactic medium (which may be pre-enriched at a low
level) or via discrete
subunits.  However, the high metallicity observed in the halos of 
hot gas around local ellipticals \citep{HB05} suggests that this
is not the dominant mechanism to form the diffuse X-ray halo.
Re-acquiring metal-enriched wind material allows for the growth of 
the X-ray halo while also maintaining a high metallicity.  It is
also possible that the enriched wind material will mix with the
ambient intergalactic medium, thereby diluting its metallicity.
Effects such as this may be responsible for the scatter in the
observed metallicities of X-ray halos or in radial metallicity
gradients.

\subsubsection{Stellar Mass Loss}

Owing to the large population of stars formed during the merger, we
expect a significant contribution to the hot gas content from ongoing
stellar mass loss.  The generation of diffuse hot gas from stellar
mass loss has been discussed extensively in the literature 
\citep[see, e.g.,][]{Cio91} and is generally punctuated by a brief period
of outflows triggered by ongoing Type I supernovae and stellar mass loss,
followed by an extended period ($\sim10$~Gyr) in which gas is retained,
and the hot gaseous halo then builds up.
We emphasize that this ``wind'' is in addition to and 
would follow the galactic wind induced by BH feedback.
This long timescale for the build up of the X-ray halo is consistent
with the observations of elliptical galaxies and merger remnants,
as discussed by \citet{OFP01}.  While our merger simulations do not alter
this picture, they do provide a physically motivated initial condition
that includes the mass of young stars.

For the simulations we study here, the majority of the 
gas initially present in the progenitor galaxies is consumed by
star formation.  Thus the mass of young stars is typically 3
- 4 times that of the hot gas, which is either in the X-ray
halo or unbound.  While the
amount of mass that is recycled from new stars back into the 
interstellar medium depends
on the assumed initial mass function; if, for example, we 
assume a value of $\sim25$\%, then stellar mass loss will
contribute an amount of hot gas that is roughly
equivalent to what already exists from the merger alone.

\subsection{Galactic Winds}
\label{ssec:winds}

A generic feature of the merger simulations that include feedback
from an accreting black hole
is the production of a large-scale outflow of metal-enriched gas.
This outflow carries a significant fraction of the available 
gas mass and eventually deposits this gas throughout the halo and
in some cases out of the galactic potential.
Winds such as these may play a significant role in
redistributing metals and may be required in order to terminate
star formation and produce red ellipticals \citep{SdMH05red,Hop05Red}.

Observationally, there is evidence that winds such as these 
exist \citep[for a review see][]{VDB05}.  At low redshift, 
winds are mainly observed in starburst galaxies via metal-line
absorption and emission studies.  It has been shown that there is
a correlation between the outflow rate and the star formation rate
and that some of these winds are powerful enough to escape the
galaxy \citep{Ru05agn,Ru05sb,M05}.  At high redshift, evidence
for large-scale outflows comes from blueshifted absorption lines,
redshifted Ly$\alpha$ emission lines \citep{Pet02,Adel03,Shap03},
and the excess of CIV absorption systems and a deficit of neutral
hydrogen near Lyman break galaxies \citep{Adel05}.

It is currently unclear how closely the winds produced in the 
simulations resemble observed winds.  We note that understanding
these winds in more detail may require embedding our simulations
in a diffuse intergalactic medium, as many of the observations
are actually a signature of the interaction between the wind and
the local diffuse medium.  This fact should be kept in mind when
interpreting the fractional
wind material presented in Figure~\ref{fig:scalings}.  Theoretical
models predict that all galaxy halos, including those of spirals,
should contain some hot gas component, yet it is currently
unclear if these hot gas halos are observed \citep{Tof02,Ben00}.  In any
case, including diffuse gas will provide a backstop against
which the outflowing gas will interact, causing it to
thermalize some of its energy, slow down, and possibly decrease
the amount of outflowing material.
We have performed one simulation in which both initial disks
included a diffuse
gaseous halo with one-quarter the baryonic mass.  In this
experiment the wind punched through the diffuse gas, and this resulted
in very little difference from the case without any diffuse gas.
More work will be necessary to determine how the wind
interacts with diffuse gas and to compare these simulations to the
observations.

As a final point, we note that the absence of these outflows in
simulations without black holes does not mean that winds are only
produced by accreting black hole feedback.  This is quite apparent
observationally, where systems such as M82 contain significant outflows
and a corresponding starburst, but there is no evidence for an
accreting black hole.  In our simulations the outflows are produced by 
point-like energy injection and one could easily consider 
other formulations of star formation \citep[see e.g.,][]{SH03}
that would also produce galactic outflows.

\section{Summary and Conclusions}
\label{sec:conc}

Our hydrodynamic simulations show that collisions between gas-rich
disk galaxies generate appreciable X-ray emission.  This emission
comes from hot diffuse gas that is produced by strong shocks
attending the merger process.  The presence of an accreting black
hole increases the X-ray luminosity during the merger through the
production of a hot, metal-rich galactic wind.  This galactic wind
transports a significant fraction of metals from the galactic center
to the diffuse gaseous halo and outside the potential well of
the remnant galaxy.

The wind-induced metal enrichment appears to be necessary to 
reproduce the near solar metallicity of observed ellipticals
\citep{HB05}.  Transporting metals
out of the galaxy potential well may also help explain the high
metallicity of the intra-group and intra-cluster media and the
apparent deficit of baryons in lower mass galaxies.

Our merger remnants span a wide range of luminosities, yet are broadly
consistent with X-ray properties of elliptical galaxies.
This result suggests that galaxy mergers are a viable mechanism for
generating the observed hot gas present in most interacting and
merging systems.  In addition, we argue that both stellar mass loss
and the reacquisition of galactic wind material are viable mechanisms
to increase the luminosity of the diffuse X-ray halo.

This paper shows that galaxy mergers may play an integral role
in the production of the hot, metal-rich gaseous halos that are 
present in most
luminous elliptical galaxies.  Using simulations similar to ours,
recent studies demonstrate that gas-rich galaxy mergers
that include black hole accretion and feedback
can reproduce the black hole
mass- central velocity dispersion relation \citep{dMSH05}
and the colors \citep{SdMH05red}, 
fundamental plane \citep{Rob05fp}, and
kinematics of elliptical galaxies \citep{Cox05rot}.
Together,
these results provide strong evidence for a ``cosmic cycle''
\citep[see Fig. 1 of][]{Hop05big}
of galaxy formation and evolution, in which gas-rich galaxy
mergers and the growth of supermassive black holes play key roles.

\acknowledgments We thank Ewan O'Sullivan, Duncan Forbes, and Jesper
Sommer-Larsen for useful
comments, and the referee for suggestions that significantly improved
this work.  This work was supported in part by NSF grants ACI
96-19019, AST 00-71019, AST 02-06299, and AST 03-07690, NASA ATP
grants NAG5-12140, NAG5-13292, and NAG5-13381.  The simulations were
performed at the Center for Parallel Astrophysical Computing at the
Harvard-Smithsonian Center for Astrophysics.

\end{document}